\documentclass[twocolumn,usenames,dvipsnames]{aastex631}

\usepackage{graphicx}
\usepackage{bm}
\usepackage{amssymb,amsmath}
\usepackage{mathrsfs} 
\usepackage{latexsym}
\usepackage{mathtools} 
\usepackage{color}
\usepackage[normalem]{ulem} 
\usepackage{comment}
\usepackage{tikz}
\usepackage{hyperref}
\usepackage{natbib}
\usepackage{multirow}
\usepackage{xspace}
\usepackage[nolist,nohyperlinks]{acronym}

\usepackage[deletedmarkup=xout,commentmarkup=uwave,commandnameprefix=ifneeded]{changes}
\renewcommand{\replaced}{\chreplaced}
\renewcommand{\deleted}{\chdeleted}

\definechangesauthor[name=Max, color=teal]{MI}
\definechangesauthor[name=Max, color=purple]{AZ}
\definechangesauthor[name=Max, color=brown]{AH}

\allowdisplaybreaks

\graphicspath{{../figures/}}

\newcommand{\Austin}{\affiliation{Weinberg Institute, University of Texas at Austin, Austin, TX 78712, USA}}
\newcommand{\CCA}{\affiliation{Center for Computational Astrophysics, Flatiron Institute, NY}}

\newcommand{\R}[1]{\textcolor{red}{#1}}

\newcommand{\rapidspinfraction}{\ensuremath{0.2^{+0.18}_{-0.18}}\xspace}
\newcommand{\rapidspinpercent}{\ensuremath{20^{+18}_{-18} \%}\xspace}

\newcommand{\rapidspinfractionleaveoneout}{\ensuremath{0.15^{+0.22}_{-0.15}}\xspace}

\newcommand{\zerospinBFDominant}{\ensuremath{7^{+71}_{-4}}\xspace}
\newcommand{\zerospinBFSubDominant}{\ensuremath{6^{+4}_{-3}}\xspace}

\newcommand{\allsecondarieszeroBF}{\ensuremath{10^{+6}_{-3}}\xspace}
\newcommand{\allsecondarieszeroBFoveronecomponent}{\ensuremath{13^{+3}_{-2}}\xspace}
\newcommand{\dominantsecondarieszeroBF}{\ensuremath{15^{+2}_{-2}}\xspace}

\newcommand{\dominantcorrelatedBF}{\ensuremath{2.9^{+0.1}_{-0.1}}\xspace}

\newcommand{\bothcorrelatedBF}{\ensuremath{2.5^{+0.2}_{-0.2}}\xspace}

\newcommand{\onepopuncorrelatedBF}{\ensuremath{2.2^{+0.1}_{-0.1}}\xspace}
\newcommand{\onepopuncorrelatedIIDBF}{\ensuremath{1.3^{+0.2}_{-0.3}}\xspace}

\newcommand{\BFsecondcomponentexists}{2.1^{+0.3}_{-0.3}}
\newcommand{\BFleaveoneoutagainstsecondcomponentexists}{1.4^{+0.1}_{-0.1}}
\newcommand{\BFslowlyspinningtwocomponentdominant}{29^{+12}_{-12}}

\begin{document}

\begin{acronym}
    \acro{GW}{gravitational wave}
    \acro{GR}{general relativity}
    \acro{CBC}{compact binary coalescence}
    \acro{BH}{black hole}
    \acro{BBH}{binary black hole}
    \acro{LVK}{LIGO-Virgo-KAGRA}
    \acro{PE}{parameter estimation}
    \acro{FAR}{false-alarm rate}
    \acro{GWOSC}{the Gravitational Wave Open Science Center}
    \acro{SSB}{solar system barycenter}
    \acro{SPA}{stationary phase approximation}
    \acro{PN}{post-Newtonian}
    \acro{BNS}{binary neutron star}
    \acro{IMR}{inspiral-merger-ringdown}
    \acro{NSF}{National Science Foundation}
    \acro{GMM}{Gaussian mixture model}
    \acro{MCMC}{Markov chain Monte Carlo}
    \acro{BIC}{Bayesian information criterion}
	\acro{BF}{Bayes factor}
	\acro{SDDR}{Savage Dickey density ratio}
	\acro{TGMM}{truncated Gaussian mixture model}
	\acro{PPD}{population predictive distribution}
	\acro{HPDI}{highest posterior density interval}
	\acro{AGN}{active galactic nucleus}
	\acro{KDE}{kernel density estimate}
\end{acronym}

\title{Hints of spin-magnitude correlations and a rapidly spinning subpopulation of binary black holes}

\author{Asad Hussain}
\Austin
\author{Maximiliano Isi}
\CCA
\author{Aaron Zimmerman}
\Austin
\date{\today}

\begin{abstract}
The complex astrophysical processes leading to the formation of binary black holes and their eventual merger are imprinted on the spins of the individual black holes.
We revisit the astrophysical distribution of those spins based on gravitational waves from the third gravitational wave transient catalog \citep[GWTC-3,][]{KAGRA:2021vkt},
looking for structure in the two-dimensional space defined by the dimensionless spin magnitudes of the heavier ($\chi_1$) and lighter ($\chi_2$) component black holes.
We find support for two distinct subpopulations with greater than 95\% credibility.
The dominant population is made up of black holes with small spins, preferring $\chi_1 \approx 0.2$ for the primary and $\chi_2 \approx 0$ for the secondary; we report signs of an anticorrelation between $\chi_1$ and $\chi_2$, as well as as evidence against a subpopulation of binaries in which both components are nonspinning.
The subdominant population consists of systems in which both black holes have relatively high spins and contains \rapidspinpercent of the binaries.
The binaries that are most likely to belong in this subpopulation are massive and slightly more likely to have spin-orientations aligned with the orbital angular momentum---potentially consistent with isolated binary formation channels capable of producing large spins, like chemically homogeneous evolution.
This hint of a rapidly spinning subpopulation hinges on GW190517, a binary with large and well-measured spins.
Our results, which are enabled by novel hierarchical inference methods, represent a first step towards more descriptive population models for black hole spins, and will be strengthened or refuted by the large number of gravitational wave detections expected in the next several years.
\end{abstract}

\section{Introduction}

\Ac{GW} detections by the \ac{LVK} Collaboration \citep{LIGOScientific:2014pky,VIRGO:2014yos,KAGRA:2020tym}
have opened a unique window onto compact objects like \acp{BH} and neutron stars, as well as the massive stars that produce them.
In particular, the vast majority of \ac{GW} detections are of \acp{BBH} 
\citep{LIGOScientific:2018mvr,LIGOScientific:2020ibl,LIGOScientific:2021usb,KAGRA:2021vkt,Nitz:2018imz,Nitz:2020oeq,Nitz:2021uxj,Nitz:2021zwj,Zackay:2019tzo,Venumadhav:2019tad,Venumadhav:2019lyq,Zackay:2019btq,Olsen:2022pin,Mehta:2023zlk}, 
which are otherwise invisible.

The distribution of spins of the individual \acp{BH} in these binaries may hold clues about their origin, e.g., whether they evolve from an isolated stellar binary or they are dynamically formed in dense environments \citep[see, e.g., reviews by][]{Mapelli:2018uds,Mandel:2018hfr}.
The dimensionless spin magnitudes, in particular, may reveal how angular momentum is distributed in the stellar progenitors and captured by the \acp{BH} at birth, as well as carry imprints of binary interactions after the first \ac{BH} forms \citep{Belczynski:2017gds,Qin:2018vaa,Fuller:2019ckz,Fuller:2019sxi,Ma:2019cpr,Bavera:2020inc,2021A&A...647A.153B,2021PhRvD.103f3032S,Zevin:2022wrw,Marchant:2023ncp}.
Spin magnitudes may additionally identify hierarchical \acp{BBH}, whose component \acp{BH} are themselves the product of previous mergers \citep{2017PhRvD..95l4046G,2019PhRvD.100d3027R,2020ApJ...900..177K,2021ApJ...915L..35K,2020ApJ...893...35D,2021NatAs...5..749G,2024MNRAS.531.3479M,Payne:2024ywe}.

Past studies of \ac{LVK} data have explored the distribution of \ac{BH} spins under different, more or less restrictive, assumptions.
Since measuring individual component spins can be difficult \citep{vanderSluys:2007st,Raymond:2009cv,Cho:2012ed,OShaughnessy:2014shr,Vitale:2014mka,Ghosh:2015jra,Chatziioannou:2018wqx,Pratten:2020igi,Green:2020ptm,Biscoveanu:2020are,Biscoveanu:2021nvg,Varma:2021csh,Miller:2023ncs,Miller:2024sui}, many analyses have looked at derived quantities like the effective spin $\chi_{\rm eff}$, which is a mass-weighted average of the spin components along the orbital angular momentum \citep{Damour:2001,Ajith:2009bn}, finding that this quantity must be small but likely positive in most systems \citep[e.g.,][]{LIGOScientific:2020kqk,KAGRA:2021duu,KAGRA:2021duu,Miller:2020zox,Callister:2021fpo,Roulet:2021hcu,Adamcewicz:2022hce,Biscoveanu:2022qac,Franciolini:2022iaa, Garciabellido:2021}.

Other works have directly tackled the individual spin magnitudes $\chi_i$ of the heavier ($i=1$) and lighter ($i=2$) components of the binary, typically assuming that they are independently and identically drawn from a unimodal distribution~\citep{2018PhRvD..97d3014W,LIGOScientific:2018jsj,LIGOScientific:2020kqk,KAGRA:2021duu}; those measurements constrain spin magnitudes to be small, $\chi_i \approx 0.2$, but with wide uncertainties.
Motivated by predictions like \cite{Fuller:2019sxi}, such models have been enhanced to look for a subpopulation of nonspinning \acp{BBH}:
while earlier studies found evidence of two populations, one with negligibly small spins and the other with larger spins \citep{Galaudage2021,Roulet:2021hcu,2022ApJ...928...75H,Kimball:2020qyd}, reanalyses with more events show no clear evidence for or against it \citep{Tong2022,Callister:2022qwb}.
Finally, a few studies have modeled the spins of the primary and secondary objects as drawn from distinct, independent distributions \citep{Tong2022,Mould:2022xeu,Adamcewicz2024,Golomb:2022bon,Edelman:2022ydv};
these measurements agree that the component spins have typical values ${\sim}0.2$ with a wide spread, and \cite{Mould:2022xeu} finds hints that the secondary could tend to have lower spins.

In this paper, we take another look at the population of \ac{BH} spin magnitudes, this time studying the structure in the joint distribution of the component spins.
Our main motivation is to look for features in the two-dimensional $\chi_1{-}\chi_2$ plane that may have escaped previous analyses because of their assumption of independent components: 
information about $\chi_1{-}\chi_2$ correlations is destroyed, and evidence of subdominant populations may be washed away, when the spins are treated as independent.
Additionally, we implement a novel technical framework that allows us to model arbitrarily narrow features in the population and treat boundary effects in the spin magnitude domain without bias.
This allows us to revisit the existence of a subpopulation of nonspinning \acp{BH} while overcoming some of the technical hurdles that have challenged previous studies.

In what follows, we describe our population model and dataset in Sec.~\ref{sec:Methods}, our population inferences in Sec.~\ref{sec:Results}, and the astrophysical implications of our results in Sec.~\ref{sec:Implications}.
We discuss our conclusions and future prospects in Sec.~\ref{sec:Conclusions}.
Additional details on our methods are given in Appendix~\ref{sec:MethodsAppx} and further results are given in Appendix~\ref{sec:ResultsAppx}. 
More details about methodology for \ac{BBH} population inference, related methods, and additional applications are described in a companion paper \cite{TGMM_Methods}.

\section{Methods and population models}
\label{sec:Methods}

We use hierarchical Bayesian inference to infer the population properties of \acp{BBH} ~\citep[e.g.,][]{Loredo:2004nn,Mandel:2018mve,Thrane:2018qnx,Vitale:2020aaz}.
The goal is to compute posteriors over the hyper-parameters $\boldsymbol{\Lambda}$ of our chosen population model.
In this study we adopt a flexible, two-population model for the BBH spin magnitudes, $\bm{\chi}=(\chi_1, \chi_2)$, drawing them from a mixture of two correlated and truncated 2D Gaussians (indexed by $a$ and $b$) with a mixing fraction $\eta$,
\begin{equation}
	p(\bm{\chi}) = \eta N_{[\boldsymbol{0},\boldsymbol{1}]}
	\left(\bm{\chi} \mid \bm{\mu}^a, \boldsymbol{\Sigma}^a\right)
	+(1-\eta) N_{[\boldsymbol{0},\boldsymbol{1}]}
	\left(\bm{\chi} \mid \bm{\mu}^b, \boldsymbol{\Sigma}^b\right),
	\label{eq:twocomponentmodel}
\end{equation}
where the $[\boldsymbol{0},\boldsymbol{1}]$ subscript indicates truncation of our domain to the $[0,1]\times[0,1]$ unit square, while both $\bm{\Sigma}^a$ and $\bm{\Sigma}^b$ independently have the general form
\begin{equation}
	\bm{\Sigma}  = \begin{pmatrix}
		\sigma_1^2 & \rho \sigma_1 \sigma_2 \\
		\rho \sigma_1 \sigma_2 &  \sigma_2^2
	\end{pmatrix} .
\end{equation}
Since we have two identical Gaussians, we face a label-switching degeneracy \cite[e.g.][]{2019PhRvD.100h4041B}, which we break by assigning an identity to the dominant population ($a$), requiring $\eta \in [0.5,1]$. 
We use truncated Gaussians rather than the Beta distributions used in \cite{KAGRA:2021duu} to more easily generalize to two dimensions and to better represent the edges of our truncated domain without systematics.
Aside from the use of truncated Gaussians, Eq.~\eqref{eq:twocomponentmodel} encompasses a wide variety of spin magnitude models used in previous studies.

\begin{figure*}[tb]
	\centering
	\includegraphics[width = 0.44 \textwidth{}]{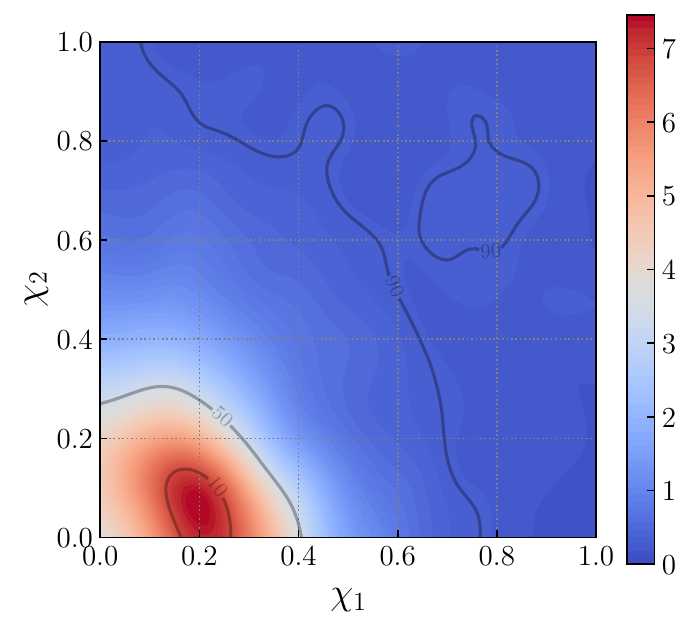}
	\includegraphics[width = 0.48 \textwidth{}]{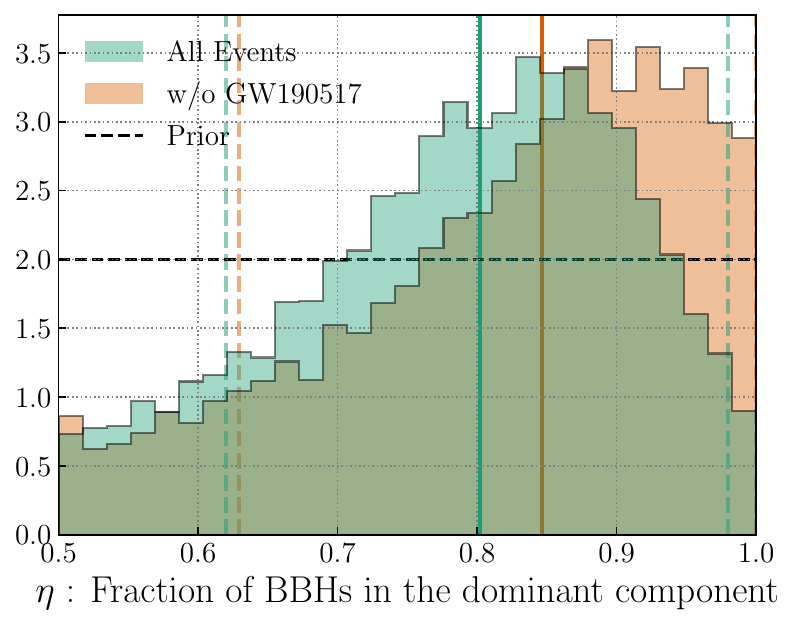}
	\caption{{\it Left:} \ac{PPD} of the spin magnitudes in our two-component model. 
	A hint of the subdominant mode is visible at  at high $\chi_1$ and $\chi_2$. 
	This mode contains \rapidspinpercent of the \acp{BBH} and is diffuse, hence is not very apparent in the \ac{PPD} despite its statistical significance. 
	The dominant, slowly spinning mode shows hints of the anticorrelation between the primary and secondary spin magnitudes.
	{\it Right:} Marginalized posterior over the fraction of \acp{BBH} in the dominant, slowly spinning mode ($\eta$). We also show the median and 90\% \ac{HPDI}. When all events are included the data prefers the existence of the highly spinning mode ($\eta \to 1$ is disfavored). However, this subpopulation is disfavored when GW190517 is removed.}
	\label{fig:TwoComponentPosteriorPredictive}
\end{figure*}

For the remaining \ac{BBH} parameters, we use the fiducial mass and redshift models defined in \cite{KAGRA:2021duu}, and the fiducial model for the tilt angles of the spins with respect to the orbital angular momentum, which assumes both spins are drawn from a two-population model: one uniform in tilt angle (isotropic spins) and one drawn from a half-normal peaking at aligned spins.
Together with Eq.~\eqref{eq:twocomponentmodel}, this makes up our full population likelihood $\mathcal L(\{d_i\} \mid \boldsymbol{\Lambda})$, where $d_i$ represents the data for the $i$th detection; with a population prior $p(\boldsymbol{\Lambda})$ and an estimate of the detection efficiency $\xi(\boldsymbol{\Lambda})$, we can then compute the population posterior $p(\boldsymbol{\Lambda} \mid \{d_i\})$ while accounting for selection effects \citep{Mandel:2018mve,Loredo:2004nn}.

Population inference requires estimating high-dimensional integrals, which
can be challenging for standard Monte Carlo methods \citep[e.g.,][]{Farr:2019rap} when population features are narrow or concentrated at the edges of the domain.
This is the case when looking for a subpopulation of \acp{BBH} with negligible spins ($\chi_i \to 0$).
To accurately compute these integrals, we first represent both single-event posteriors and detection-probability estimates as \acp{TGMM}.
This allows us to leverage properties of Gaussians to analytically evaluate integrals over the spin sector (magnitudes and tilts), while using Monte Carlo averages over the remaining parameters.
Our methods help to control the variance in the estimates of our likelihood integrals across hyper-parameter space, and so we do not apply data-dependent priors to exclude regions of high variance \cite[unlike, e.g.,][]{KAGRA:2021duu}.
We discuss our strategy in detail in a companion paper \citep{TGMM_Methods}, and summarize it in App.~\ref{sec:MethodsAppx}.

We visualize the result of our fits by plotting the \ac{PPD} for the spin magnitudes, which represents the inferred distribution of spins marginalized over all our population parameters, i.e.,
\begin{equation}
	p(\bm{\chi} \mid \{d_i\}) = \int p(\bm{\chi} \mid \boldsymbol{\Lambda})\, p(\boldsymbol{\Lambda} \mid \{d_i\}) \, \mathrm{d}\boldsymbol{\Lambda}\, .
	\label{eq:PPD}
\end{equation}
We also show projections of the population posterior for different hyper-parameters.
Additionally, to perform model comparison of our fiducial spin-magnitude model against lower-dimensional subcases, we use the \ac{SDDR} for nested models \citep{dickey1971weighted}.
To compute this, we use tailored methods to construct unbiased truncated \acp{KDE} of our population posteriors to evaluate them at the limiting points~\citep{TGMM_Methods},
and bootstrap over multiple hyper-posterior draws to report a median estimate and a 90\%-confidence highest-density interval (see also Appendix~\ref{sec:MethodsAppx}).
Unbiased density estimation on boundaries is a well-known challenge in many settings, and our methods have benefits over standard solutions like reflective \acp{KDE}, for example not imposing a zero derivative at the boundary (see also Appendix~\ref{sec:MethodsAppx}).

For our dataset we use the 69 confidently detected (false alarm rates below $1/\mathrm{yr}$) events used by~\cite{KAGRA:2021duu} for \ac{BBH} population inferences.
We use posterior samples produced using the \textsc{IMRPhenomXPHM} waveform model~\citep{Pratten:2020ceb} released a part of the GWTC-2.1 and GWTC-3 catalogs~\citep{LIGOScientific:2021usb,KAGRA:2021vkt}, and available as open data~\citep{LIGOScientific:2019lzm,KAGRA:2023pio} at \cite{GWOSC}. 
To incorporate selection effects, we use the sensitivity estimates described in~\cite{KAGRA:2021duu} and provided by~\cite{ligo_scientific_collaboration_and_virgo_2021_5636816}.
We sample our hyper-posteriors using the no-U-turn sampler \citep{2011arXiv1111.4246H} Hamiltonian Monte Carlo \citep{2011hmcm.book..113N,2017arXiv170102434B} implemented in \textsc{numpyro}~\citep{phan2019composable,bingham2019pyro}.

\section{Results}
\label{sec:Results}

We present our main result in Fig.~\ref{fig:TwoComponentPosteriorPredictive}, showing the \ac{PPD} over $\chi_1$ and $\chi_2$ (left) and the posterior on the mixture fraction $\eta$ (right).
The \ac{PPD} reveals a bimodal distribution of spins, with a dominant component peaking at $\chi_1 \approx 0.2$ and $\chi_2 \approx 0$ and a subdominant component peaking at $\chi_i \approx 0.75$.
In Fig.~\ref{fig:TwoComponentPosteriorPredictiveByComponent}, we isolate the contributions of each component, making it clear that the dominant mode consists of low spins \acp{BH} (top), while the subdominant mode mostly supports high spins (bottom). While the dominant mode is quite well measured, the subdominant mode is localized more diffusely, as might be expected given its lower occupancy.
The subdominant component makes a relatively small contribution to the \ac{PPD} in Fig.~\ref{fig:TwoComponentPosteriorPredictive}, but this is a function of the smaller number of events that are assigned to this mode and not a measure of our certainty in its existence.

To asses the significance of the second mode, the right panel of Fig.~\ref{fig:TwoComponentPosteriorPredictive} shows our inferred posterior over the fraction of systems in the dominant subpopulation.
The fraction of \acp{BBH} in the subdominant component is $1-\eta=$ \rapidspinfraction quoting the \ac{HPDI} around the median. 
In other words, ${\sim}20\%$ (${\sim}80\%$) of the \acp{BBH} are favored to be in the rapidly (slowly) spinning subpopulation.
The posterior on $\eta$ rules out a single population of spin-magnitudes ($\eta = 1$) at better than $95\%$ credibility.
The \ac{BF} in favor of the two-component model is $\mathcal{B}(\text{2 vs 1}) = \BFsecondcomponentexists$ (see Sec.~\ref{sec:Methods} and Appendix~\ref{sec:SDDRMethods}).

The evidence for the subdominant mode is highly sensitive to the rapidly spinning event GW190517 \citep{LIGOScientific:2020ibl}.
Removing it from our set, the data no longer support the a subpopulation, with $1-\eta = \rapidspinfractionleaveoneout$ encompassing zero within $90\%$ credibility,
and instead yielding a \ac{BF} against the existence of this subpopulation of $\mathcal{B}(\text{1 vs 2}) \approx \BFleaveoneoutagainstsecondcomponentexists$. 
The source of GW190517 is decisive due to the fact that its spin magnitudes are confidently measured to be large (see, e.g., Fig.~10 of \citealt{LIGOScientific:2020ibl} or Fig.~2 of \citealt{Qin:2022ncz}), and cannot be accommodated by the dominant population alone.
On the other hand, our results are insensitive to the exclusion of other GW events from highly spinning \acp{BBH}, such as GW191109 \citep{KAGRA:2021vkt}.
No data quality issues have been reported for GW190517.

Although previous works have reported hints of a potential subpopulation of rapidly spinning \acp{BH} \citep{Galaudage2021,Roulet:2021hcu,2022ApJ...928...75H}, it would be difficult for such studies to clearly identify the secondary mode in Fig.~\ref{fig:TwoComponentPosteriorPredictive} because such a subpopulation is not apparent in the $\chi_1$ or $\chi_2$ marginals (see Fig.~\ref{fig:PPD_marginals} in Appendix~\ref{sec:ResultsAppx}), as it is obfuscated by the tails of the low-spin mode.
On the other hand, the subpopulation stands out more clearly in 2D because those high-spin systems would otherwise have to be accommodated by the tail of the dominant component in both $\chi_1$ and $\chi_2$ \emph{simultaneously}, which is made difficult by the fact that the majority of events constrain the bulk of the posterior to be in a compact ball near the origin (in other words, the secondary mode lies well beyond the 90\% credibility contour in 2D, but not in the 1D marginals).

When comparing to studies that allow for a subpopulation with negligible spins, we must look at two cases, one where each of our Gaussian components separately concentrates at the origin.
Using the \ac{SDDR}, we find a \ac{BF} against a dominant population with negligible spin of $\mathcal B = \zerospinBFDominant$
and a \ac{BF} against a subdominant population with negligible spins $\mathcal B = \zerospinBFSubDominant$.
As compared to our other \ac{SDDR}-based \ac{BF} comparisons, these are especially uncertain because we must extrapolate our hyper-posterior samples to a corner in 4D space ($\mu_1^a = \sigma_1^a = \mu_2^a = \sigma_2^a = 0$), but in both cases we clearly disfavor a subpopulation with negligible spins.

We next describe the features of each subpopulation in more detail.
Additional corner-plots of the hyper-posteriors are shown in Appendix~\ref{sec:ResultsAppx}.

\subsection{The dominant population: slow and anticorrelated spins}

\begin{figure}[tb]
\includegraphics[width = 0.5 \textwidth{}]{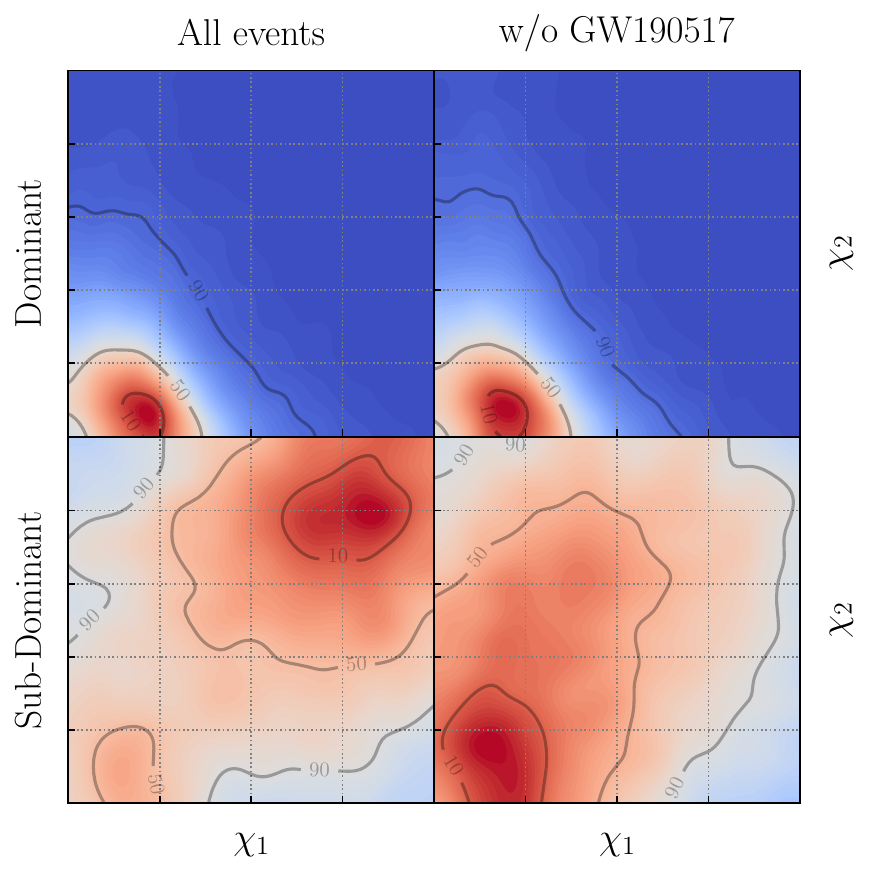}
	\caption{Separate \acp{PPD} of the spin magnitudes of the dominant and subdominant populations, shown with and without the inclusion of GW190517. 
	The subdominant mode prefers higher spins when GW190517 is included. 
	Without GW190517 the peak moves to lower spin values and is degenerate with the dominant mode, indicating that the data disfavor a subpopulation.
 The color scale normalization varies from panel to panel to resolve the variations in the \ac{PPD}.}
	\label{fig:TwoComponentPosteriorPredictiveByComponent}
\end{figure}

\begin{figure}[tb]
	\includegraphics[width = 0.5 \textwidth{}]{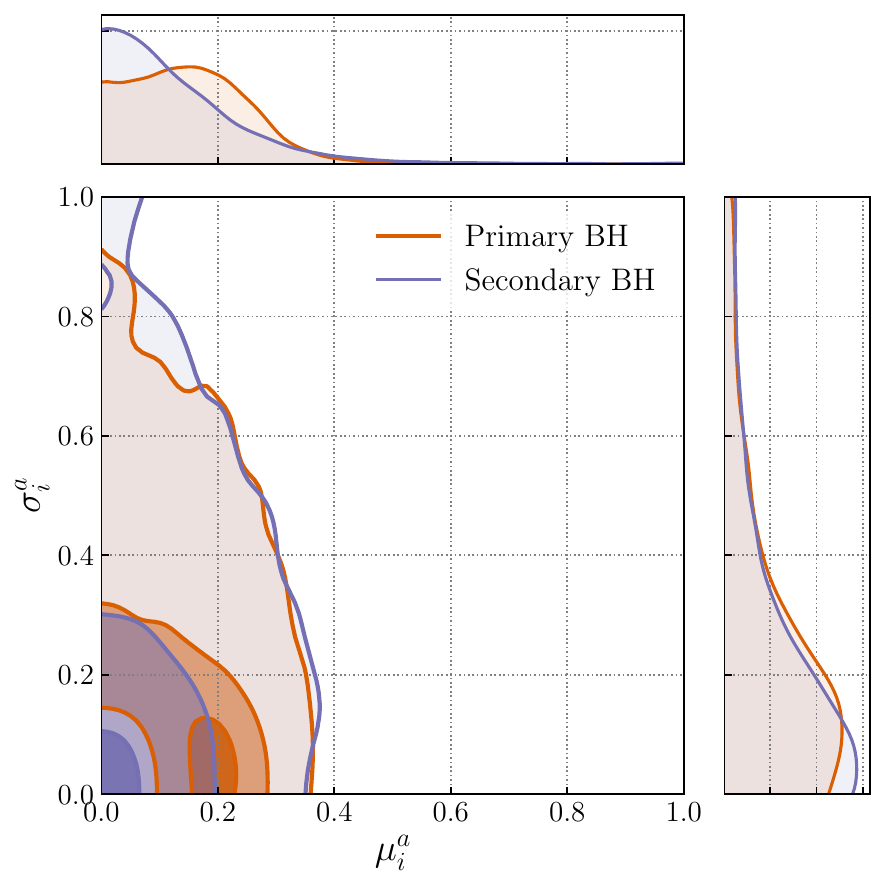}
	\caption{Dominant subpopulation. Posteriors for the mean ($\mu^{a}_i$) and scale ($\sigma^{a}_i$) population parameters of the primary BH (orange) and secondary BH (purple) in the dominant subpopulation with $90\%, 50\%$ and $10\%$ credible intervals marked. 
	The data prefer that for the secondaries $\mu^a_2 \to 0$, $\sigma^a_2 \to 0$ (i.e.~all secondary \acp{BH} nonspinning). 
	In addition there is a degeneracy between $\mu^a_1$ and $\sigma^a_1$ for the primary's spin. 
	The data support a sharp peak with all primaries having $\chi_1 \approx 0.2$, and is also consistent with $\chi_1$ drawn from a half-normal peaking at $\chi_1 = 0$ with a spread of $\approx 0.2$}
	\label{fig:TwoComponentDominantCorner}
\end{figure}

The dominant mode in the population 
has a number of interesting features: (1) both component spins are well constrained to be low, (2) there are differences between the primary and secondary spins, and (3) there is a hint of anticorrelation between the spins.
We discuss each in turn.

The fact that \acp{BH} in this subpopulation tend to spin slowly, regardless of whether they are the lighter or heavier component in the binary, is evident from the \ac{PPD} in the top panel of Fig.~\ref{fig:TwoComponentPosteriorPredictiveByComponent}; since this mode dominates the population, it can also be gleaned from Fig.~\ref{fig:TwoComponentPosteriorPredictive}.
Additionally, the preference for low spins can be seen in the inferred population mean and scale parameters for this mode (Fig.~\ref{fig:TwoComponentDominantCorner}), which strongly favor low values for both components in the binary.
The uncertainties on $\chi_2$ are not significantly worse than $\chi_1$.
Since the dominant mode contains most \acp{BBH}, it has properties similar to those inferred in past spin population studies.

Next, we find interesting differences between $\chi_1$ and $\chi_2$.
While the spin of the primary \ac{BH} peaks at $\chi_1 \approx 0.2$, as expected from previous studies, the secondary \ac{BH} population is consistent with identically vanishing spins.
Not only does the \ac{PPD} peak at $\chi_2 \approx 0$ but also the population posterior favors a delta function at $\chi_2 = 0$ (the purple distribution peaks at $\mu_2^a = \sigma_2^a = 0$ in Fig.~\ref{fig:TwoComponentDominantCorner}), with a \ac{BF} of $\mathcal B = \dominantsecondarieszeroBF$ in favor of identically nonspinning secondaries.
Conditioned on $\eta \to 1$, (i.e., only one population exists) we still find that the data favors all of the secondaries to be nonspinning with a \ac{BF} of $\mathcal B = \allsecondarieszeroBF$.

Previous studies have looked for differences in the $\chi_1$ and $\chi_2$ distributions, assuming independence.
For example, \cite{Adamcewicz2024} compared a model where both \acp{BH} are spinning to one where either the primary, secondary, or both are nonspinning by repeating parameter estimation over the catalog of \acp{BBH} for each case;
we corroborate their result favoring the case where only the primaries are spinning, without the need for additional, computationally expensive parameter estimation.
Hints of the support for lower secondary spins can also be gleaned from Fig.~A1 in \cite{Mould:2022xeu};
on the other hand, \cite{Edelman:2022ydv} found no visible difference between $\chi_1$ and $\chi_2$ with a more flexible model.

Unlike the secondaries, the primaries cannot all be nonspinning.
This is clear from Fig.~\ref{fig:TwoComponentDominantCorner}, where $\mu_1^a$ and $\sigma_1^a$ are not allowed to simultaneously vanish (orange contours).
The distribution of $\chi_1$ most likely peaks at $\mu_1^a \approx 0.2$, as expected from Fig.~\ref{fig:TwoComponentPosteriorPredictiveByComponent}, but may peak at zero \emph{as long as the spread is sufficiently large} ($\sigma_1^a \approx 0.2$).
This means that the data show evidence of at least one event with a primary spin of $\chi_1 \approx 0.2$.

\begin{figure}[tb]
	\includegraphics[width = 0.48 \textwidth{}]{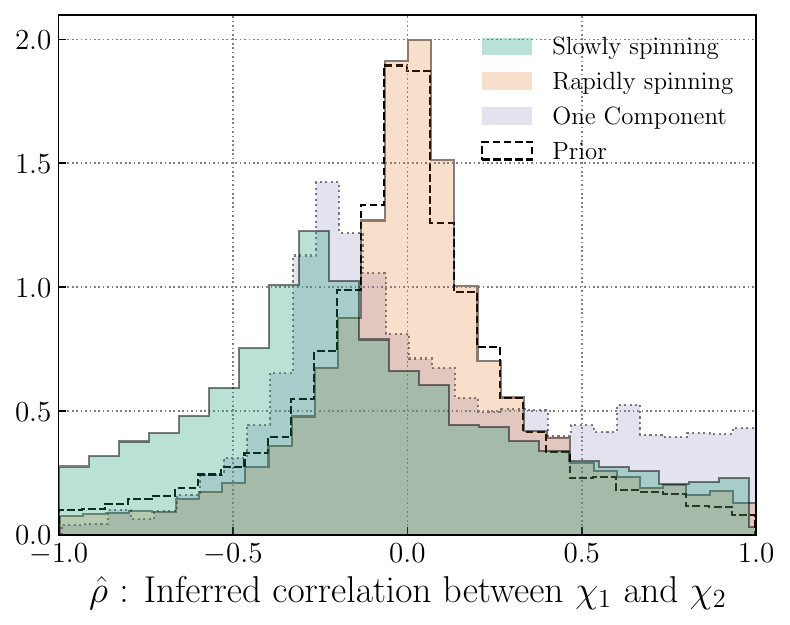}
		\caption{Posterior over the Pearson correlation coefficient $\hat{\rho}$ between the spin magnitude of the primary and secondary BH for \acp{BBH} in the different subpopulations: the slowly spinning mode in a two-component fit (green), its rapidly spinning counterpart (orange), and the result of a unimodal fit (purple); we additionally show the effective prior (dashed).} 
		\label{fig:TwoComponentCorrelations}
	\end{figure}

Finally, returning to the top panels of Fig.~\ref{fig:TwoComponentPosteriorPredictiveByComponent}, we see that 
the \ac{PPD} suggests an anticorrelation between the spin magnitudes.
While we do not rule out zero correlation ($\rho^a = 0$) at the $90\%$ credible level, the \ac{BF} against an uncorrelated distribution is $\mathcal{B} = \dominantcorrelatedBF$. 
Assessment of the correlation is complicated by the fact that a uniform prior on $\rho^a$ induces a prior on the Pearson correlation coefficient $\hat{\rho} = \text{Corr}[\chi_1, \chi_2]$ which is strongly peaked at zero.
This is because for truncated Gaussians, a large range of means, scale parameters, and $\rho$ values result in a small empirical correlation in the bounded $\chi_1$--$\chi_2$ domain.
This can be seen clearly in Fig.~\ref{fig:TwoComponentCorrelations}, where we show our posteriors and priors over  $\hat \rho$ for both subpopulations.
We see that there are hints of an anticorrelation between the spins of the dominant population, as it is able to overcome this strong prior on $\hat \rho^a$.
We also plot the case where we fix $\eta = 1$, so that we have only a single spin-magnitude population.
The peak is then similar to our fiducial model, favoring weak anticorrelation, but with a heavy tail towards positive $\hat \rho$ values.
This can be interpreted as the imprint of the highly spinning subpopulation on our single-population model, since some posterior weight is pulled towards the large $\chi_1$--$\chi_2$ region while requiring the bulk of the spin population to lie at small spins.

In any case, the correlation structure apparent in the \ac{PPD} of Fig.~\ref{fig:TwoComponentPosteriorPredictiveByComponent} suggests a preference for pairing higher spinning primaries with lower spinning secondaries and vice versa, such that systems where both \acp{BH} are nonspinning ($\chi_1= \chi_2=0$) are measurably disfavored.

\subsection{The subdominant population: relatively high spins}

The \ac{PPD} of the subdominant population (bottom left of  Fig.~\ref{fig:TwoComponentPosteriorPredictiveByComponent}) peaks at large spins, and slightly disfavors cases where one of the two \acp{BH} is rapidly spinning while the other has negligible spin.
However, the subpopulation is broad, and the posteriors on the hyper-parameters are weakly informed by the data.
Its empirical correlation $\hat \rho^b$ is consistent with the prior, as seen in Fig.~\ref{fig:TwoComponentCorrelations}.
Further, we see some contamination at small spin-magnitudes in the \ac{PPD} from the remaining degeneracy between the dominant and subdominant components, which occurs when $\eta \to 0.5$.
The bottom right plot in Fig.~\ref{fig:TwoComponentPosteriorPredictiveByComponent} shows that once we remove the highly spinning event GW190517, this subpopulation becomes degenerate with the dominant component in terms of its location but also captures the tail of the distribution of spin magnitudes towards higher values of $\chi_i$.

\begin{figure}[tb]
	\includegraphics[width = 0.48\textwidth]{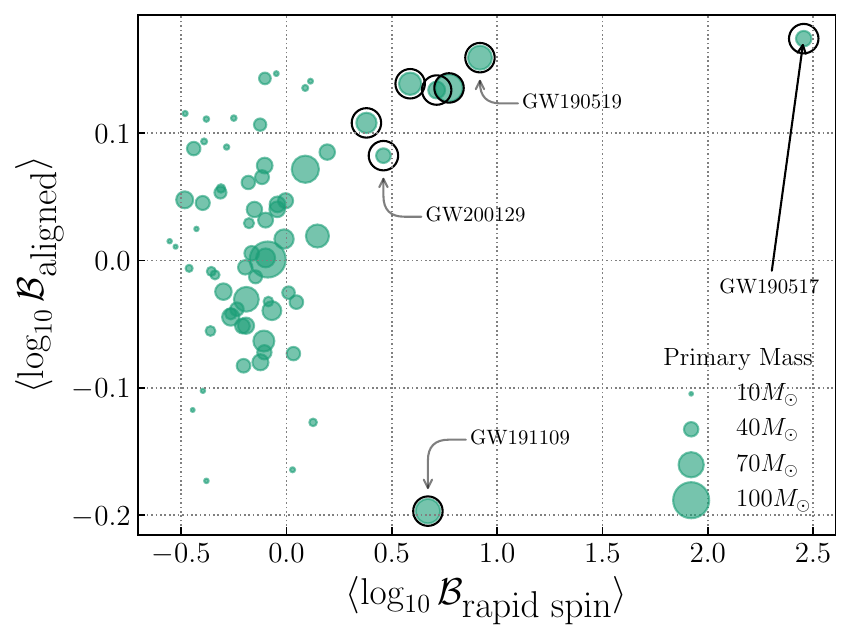}
	\caption{
		Log \acp{BF} for all 69 events, comparing the hypotheses that each \ac{BBH} comes from the rapidly spinning subpopulation versus the dominant slowly spinning population (abscissa) and that each comes from the predominantly aligned-spin subpopulation versus the isotropic subpopulation (ordinate).
		These \acp{BF} are marginalized over our hyper-posterior samples in the manner discussed in Appendix~\ref{sec:Appxeventassignments}.
		Although all these \acp{BF} are weak, the trend suggests that events more likely to lie in the rapidly spinning subpopulation are more likely to come from the predominantly aligned subpopulation and tend to have higher masses.
		Events with a median \ac{BF} greater than 2 are marked with black circles, and some events of special interest are labeled. The largest $90\%$ confidence interval of $\langle \log_{10}\mathcal B_{\text{rapid spin}}\rangle$ over all events is $\pm 0.2$,  and for $\langle \log_{10}\mathcal B_{\text{aligned}}\rangle$ is $\pm 0.03$.}
	\label{fig:OrientationMagnitudeMassTrend}
\end{figure}
	
Additional clues about the origin of this subdominant mode can be found in possible correlations with other \ac{BBH} parameters, for example the masses or spin tilts.
In Fig.~\ref{fig:OrientationMagnitudeMassTrend}, we visualize the estimated log \ac{BF} for each \ac{BBH} we analyzed to lie in the rapidly spinning versus the slowly spinning population (abscissa), as well as the log \ac{BF} that to lie in the mostly-aligned-spin versus isotropic-spin tilt population (ordinate).
Although the spin-tilt \acp{BF} are very weak, it seems that the rapidly-spinning \acp{BBH} tend to also fall in the aligned-spin population, with the exception of GW191109 which is considerably anti-aligned \citep{Udall:2024ovp};
for \acp{BH} with smaller spins it is harder to determine the tilts, leading to correspondingly larger scatter in the spin-tilt \ac{BF}.
In addition to tilts, Fig.~\ref{fig:OrientationMagnitudeMassTrend} also encodes the binary primary mass (maker size) revealing that those systems most likely to lie in the rapidly spinning population are also more massive, consistent with a correlation between mass and spin previously identified in, e.g.,~\cite{Tiwari:2020otp,2022ApJ...928...75H,Franciolini:2022iaa, Callister:2021fpo, Adamcewicz:2022hce, Biscoveanu:2022qac}.

\section{Astrophysical implications}
\label{sec:Implications}

The spin magnitudes of \acp{BBH} are determined by a number of factors, ranging from binary interactions which may tidally spin up the progenitor stars, the angular momentum distribution and transport within the progenitor stellar cores during their evolution, the uncertain details of core collapse, and the effects of any fallback accretion \citep[e.g.,][and references therein]{Mandel:2018hfr}.
At face value, the bimodality that we infer in \ac{BBH} spin magnitudes suggests that multiple formation channels may be at play (already supported by the inferred distribution of spin-tilt angles, \citealt{KAGRA:2021duu}).
This could be the case if \ac{LVK} \acp{BBH} are a mixture of dynamically-formed systems and systems from isolated stellar binaries.
Alternatively, isolated binaries alone could accommodate this bimodality if, e.g., angular momentum is transported from the stellar core to the envelope leading to slowly spinning \acp{BH} \citep{Fuller:2019sxi}, except when stars are tidally spun up~\citep[e.g.][]{Kushnir:2016zee,Fuller:2022ysb,Ma:2023nrf} or are chemically homogeneous without a core-envelope structure~\citep[e.g.][]{Mandel:2015qlu,deMink:2016vkw,Marchant:2016wow}, resulting in one or both \acp{BH} with large spins (e.g., \citealt{Qin:2018sxk}, although see~\citealt{Riley:2020btf}).

With this context, we first consider interpretations for the dominant population of slowly spinning \acp{BBH}, where we find the secondary \acp{BH} have negligible spin and the primaries have spins  concentrated near $\chi_1 \approx 0.2$, 
with hints of an anticorrelation between the component spin magnitudes, and strong support against both \acp{BH} being nonspinning.
Given the wide range of angular momentum values that stellar cores can possess before collapse~\citep[e.g.,][]{Qin:2018sxk}, it is challenging from first principles to produce natal \ac{BH} spins that are not either very high or negligible, and doubly so to have two distinct cases for the primary and secondary.
As seen in Fig.~\ref{fig:TwoComponentDominantCorner}, another acceptable scenario is one where the secondary spin is negligible and the primary distribution peaks at zero with a width $\sigma^a_1 \approx 0.25$, allowing for a range of spin magnitudes for the primary.
Both cases remain challenging to explain, since a standard binary evolution scenario might leave the first born, presumably more massive, \ac{BH} with a small spin while the secondary can be spun up by tidal effects~\citep{Hotokezaka:2017esv,Zaldarriaga:2017qkw,Qin:2018vaa}.
Alternatively, \ac{BH} spins of $\chi_i \approx 0.2$ can be explained by moderately efficient angular momentum transport mechanisms that result in such natal spins \citep{Belczynski:2017gds}, or through the accretion of a portion of the highly convective envelope onto a \ac{BH} with negligible natal spin~\citep{Antoni2022}.
However, it is not clear how to explain the distinction between primary and secondary spins in these scenarios.

One resolution to the $\chi_1$ versus $\chi_2$ asymmetry would be for a mass-ratio reversal to occur prior to the formation of the first \ac{BH}, followed by spin-up of the now more massive star through tidal interactions with the first \ac{BH} \citep[e.g.,][]{Gerosa:2013laa,Olejak:2021iux,Zevin:2022wrw,Broekgaarden:2022nst}, yielding a spinning primary and a secondary with a lower spin.
The models considered by~\cite{Broekgaarden:2022nst} indicate that mass-ratio reversal can be common among detectable \acp{BBH}, but also predict some systems with spinning secondaries and a significant number of systems with negligible spins.
However, the spin-tilt distribution provides evidence that a fraction of \acp{BBH} have isotropically distributed spins~\citep{KAGRA:2021duu}. 
Since we find that the rapidly spinning subpopulation prefers more aligned spins, we expect that some of the low-spin systems arise from the isotropic spin distribution,
indicative that some \acp{BBH} form outside of the isolated binary evolution channel, 
or strong natal kicks~\citep[e.g.,][]{Callister:2020vyz}.

The highly spinning subpopulation is intriguing. 
A key feature is that both primary and secondary spins tend to be large in this population, although with a wide range of uncertainties.
We disfavor the case where the primary has a large spin $\chi_1 \approx 0.7$ while the secondary has a small spin, disfavoring a population of mergers between first and second-generation \acp{BH} in a dense stellar environment; since such mergers would be more common than mergers between two second-generation \acp{BH}~\citep[e.g.,][]{2020ApJ...900..177K}, it is unlikely that these large BH spins are produced by previous \ac{BBH} mergers.
The correlation we find between probability of being highly-spinning and probability of having relatively aligned spins for the detected \acp{BBH} further hints that the large spins may arise from binary interactions.
Together with the fact that the highly spinning population appears to be made up of more massive \acp{BH}, the homogeneous evolution of low-metallicity binaries would appear to be a reasonable scenario for this population.

\section{Conclusions}
\label{sec:Conclusions}

In this work we have investigated the astrophysical distribution of \ac{BH} spin magnitudes based on 69 \acp{BBH} detected with high significance in \ac{GW} observations by the \ac{LVK}.
Unlike previous work, we have explored the two-dimensional space of component-spin magnitudes directly, using a model that subsumes those of many previous studies and allows for two distinct populations in spin-magnitude space.
We have confirmed that the bulk of \acp{BBH} have small but non-negligible spin magnitudes, with primaries favoring $\chi_1 \approx 0.2$, while finding new evidence that secondary \acp{BH} are consistent with having identically zero spin.
We have also found evidence for a weak anticorrelation between the spin components and a we disfavor models in which the majority of \acp{BBH} have negligible spins.
The latter result was enabled in this study by novel methods to make inferences about populations with narrow features
without recourse to tailored parameter estimation for the \ac{GW} events.

We have uncovered hints of a second, subdominant spin population containing \rapidspinpercent of the \acp{BBH}, ruling out a single component with better than 95\% credibility.
This second population, although broad, peaks at large spin magnitudes for both the primary and secondary \acp{BH}.
The evidence for this subpopulation is largely driven by a single \ac{GW} event, GW190517, whose components have high and relatively well-measured spin magnitudes.
We identify the probable \ac{GW} events which arise from the rapidly spinning subpopulation, and find that these are preferentially massive.
As in previous population studies, \cite[e.g.][]{KAGRA:2021duu}, we allow for two populations of spin orientations, one isotropic and one peaking towards alignment with the orbital angular momentum.
We find that the \acp{BBH} identified with the rapidly spinning population are somewhat more likely to be in the aligned spin population,
perhaps suggesting an origin in field binaries composed of massive, rapidly rotating stars.
Note that spin measurements can be impacted by \ac{GW} modeling systematics, and future work should assess the impact of such systematic errors on the properties and significance of the subdominant population found here.

The analysis reported here is only the first allowing for multiple populations of \acp{BBH} with correlated spin-magnitudes.
In order to better constrain the possible origin of the rapidly spinning \acp{BBH}, analyses including correlations between the spin magnitudes and spin orientations, binary masses, or redshift would be of great interest.
Targeted analysis exploring the dominant, slowly spinning population would also be highly valuable, particularly those 
that address the possibility that mass ratio reversal may play a role in forming binaries with small but non-negligible 
primary spin magnitudes and secondary spins consistent with zero.

The evidence for the rapidly spinning subpopulation and its properties remains tentative.
With only 69 events in our dataset, our ability to infer fine details and isolate subpopulations is limited.
At the time of writing, however, the \ac{LVK} is in the midst of its fourth observing run, at even greater sensitivity 
than the previous campaign.
To date over 100 public alerts have been issued reporting \ac{GW} event candidates with false alarm rates less than $2/\mathrm{yr}$, with many 
more \ac{BBH} detections expected as \ac{GW} detectors reach and exceed their design sensitivities in the coming years \citep{KAGRA:2013rdx}.
This growing dataset should confirm the rapidly spinning subpopulation if it exists, allow for more detailed inferences
about the properties of \acp{BBH} in it, and, in so doing, help to uncover the origin and formation channels of merging \acp{BBH}.
\\

\noindent We thank Christopher Berry, Tom Callister, Will Farr, Maya Fishbach, Matthew Mould, Ester Ruiz-Morales, and Colm Talbot for valuable discussions.
We especially thank Simona Miller for reviewing a draft of this work and providing helpful comments.
A.H~and A.Z.~were supported by NSF Grants PHY-1912578, PHY-2207594 and PHY-2308833 over the course of this work.
The Flatiron Institute is a division of the Simons Foundation.
This work has preprint numbers LIGO-P2400522 and UTWI-33-2024.
This material is based upon work supported by NSF's LIGO Laboratory which is a major facility fully funded by the National Science Foundation.
The authors are grateful for computational resources provided by the LIGO Laboratory and supported by NSF Grants PHY-0757058 and PHY-0823459.
This research has made use of data or software obtained from the Gravitational Wave Open Science Center (gwosc.org), a service of the LIGO Scientific Collaboration, the Virgo Collaboration, and KAGRA. 
This material is based upon work supported by NSF's LIGO Laboratory which is a major facility fully funded by the National Science Foundation, as well as the Science and Technology Facilities Council (STFC) of the United Kingdom, the Max-Planck-Society (MPS), and the State of Niedersachsen/Germany for support of the construction of Advanced LIGO and construction and operation of the GEO600 detector. 
Additional support for Advanced LIGO was provided by the Australian Research Council. 
Virgo is funded, through the European Gravitational Observatory (EGO), by the French Centre National de Recherche Scientifique (CNRS), the Italian Istituto Nazionale di Fisica Nucleare (INFN) and the Dutch Nikhef, with contributions by institutions from Belgium, Germany, Greece, Hungary, Ireland, Japan, Monaco, Poland, Portugal, Spain. 
KAGRA is supported by Ministry of Education, Culture, Sports, Science and Technology (MEXT), Japan Society for the Promotion of Science (JSPS) in Japan; National Research Foundation (NRF) and Ministry of Science and ICT (MSIT) in Korea; Academia Sinica (AS) and National Science and Technology Council (NSTC) in Taiwan.

\appendix

\section{Details on methods and population models}
\label{sec:MethodsAppx}

\begin{table*}[]
	\centering
	\begin{tabular}{l c|c|c}
		Model / Sub-model & Schematic & Prior/Limits & Comments/Bayes Factor \\
		\hline
		\hline
		&&& \\[-10pt]
		\multirow{3}{*}{ {\bf Two Components}
		} &    
		\multirow{3}{*}{\resizebox{2.2cm}{2.2cm}{      \begin{tikzpicture}[x=0.75pt,y=0.75pt,yscale=-1,xscale=1]

\draw  [color={rgb, 255:red, 255; green, 0; blue, 0 }  ,draw opacity=0.58 ][fill={rgb, 255:red, 238; green, 0; blue, 0 }  ,fill opacity=0.13 ][line width=1.5]  (62.14,166.33) .. controls (77.23,147.91) and (114.14,153.19) .. (144.58,178.13) .. controls (175.02,203.07) and (187.46,238.22) .. (172.36,256.64) .. controls (157.27,275.07) and (120.36,269.78) .. (89.92,244.84) .. controls (59.48,219.9) and (47.04,184.75) .. (62.14,166.33) -- cycle ;
\draw  [draw opacity=0][fill={rgb, 255:red, 255; green, 255; blue, 255 }  ,fill opacity=1 ] (30.5,38.36) -- (295.5,38.36) -- (295.5,295.36) -- (30.5,295.36) -- cycle(256.95,76.91) -- (69.05,76.91) -- (69.05,256.81) -- (256.95,256.81) -- cycle ;
\draw   (69.05,76.91) -- (256.95,76.91) -- (256.95,256.81) -- (69.05,256.81) -- cycle ;

\draw  [color={rgb, 255:red, 30; green, 0; blue, 255 }  ,draw opacity=0.58 ][fill={rgb, 255:red, 3; green, 0; blue, 255 }  ,fill opacity=0.13 ][line width=1.5]  (184.78,165.81) .. controls (172.71,158.06) and (173.32,135.6) .. (186.14,115.67) .. controls (198.96,95.73) and (219.13,85.86) .. (231.2,93.62) .. controls (243.27,101.38) and (242.66,123.83) .. (229.84,143.76) .. controls (217.02,163.7) and (196.85,173.57) .. (184.78,165.81) -- cycle ;

\end{tikzpicture}}}
		 & $\eta \sim U(0.5,1)$ & Fraction of \acp{BBH} in the dominant component \\ 
		 & & $\mu^{a,b}_i \sim U(0,1)$ & Mean parameters of each component \\
		$p(\boldsymbol{\chi}) = \eta N_{[\boldsymbol{0},\boldsymbol{1}]}\left(\boldsymbol{\chi}\ | \boldsymbol{\mu}^a, \boldsymbol{\Sigma}^a\right)+$& & $\sigma^{a,b}_i \sim U(0,1)$ & Scale parameters of each component \\
		$\ \ \ \ \ \ \ (1-\eta) N_{[\boldsymbol{0},\boldsymbol{1}]}\left(\boldsymbol{\chi}\ | \boldsymbol{\mu}^b, \boldsymbol{\Sigma}^b\right)\,$& & $\rho^{a,b} \sim U(-1,1)$ & Correlation parameter of each component \\
		& & & \\[2pt]
		\hline
		\hline
		\multirow{2}{*}{\qquad Two uncorrelated components} &    
	\multirow{2}{*}{\resizebox{1cm}{1cm}{\tikzset{every picture/.style={line width=0.75pt}} 

\begin{tikzpicture}[x=0.75pt,y=0.75pt,yscale=-1,xscale=1]

\draw  [color={rgb, 255:red, 255; green, 0; blue, 0 }  ,draw opacity=0.58 ][fill={rgb, 255:red, 238; green, 0; blue, 0 }  ,fill opacity=0.13 ][line width=1.5]  (46,211.49) .. controls (46,187.67) and (77.9,168.36) .. (117.25,168.36) .. controls (156.6,168.36) and (188.5,187.67) .. (188.5,211.49) .. controls (188.5,235.3) and (156.6,254.61) .. (117.25,254.61) .. controls (77.9,254.61) and (46,235.3) .. (46,211.49) -- cycle ;
\draw  [draw opacity=0][fill={rgb, 255:red, 255; green, 255; blue, 255 }  ,fill opacity=1 ] (30.5,38.36) -- (295.5,38.36) -- (295.5,295.36) -- (30.5,295.36) -- cycle(256.95,76.91) -- (69.05,76.91) -- (69.05,256.81) -- (256.95,256.81) -- cycle ;
\draw   (69.05,76.91) -- (256.95,76.91) -- (256.95,256.81) -- (69.05,256.81) -- cycle ;

\draw  [color={rgb, 255:red, 30; green, 0; blue, 255 }  ,draw opacity=0.58 ][fill={rgb, 255:red, 3; green, 0; blue, 255 }  ,fill opacity=0.13 ][line width=1.5]  (207.99,86.8) .. controls (222.34,86.8) and (233.97,106.02) .. (233.97,129.72) .. controls (233.97,153.42) and (222.34,172.63) .. (207.99,172.63) .. controls (193.65,172.63) and (182.02,153.42) .. (182.02,129.72) .. controls (182.02,106.02) and (193.65,86.8) .. (207.99,86.8) -- cycle ;

\end{tikzpicture}}}
	&  \multirow{2}{*}{$\hat{\rho}^{a,b} = 0$}  & \multirow{2}{*}{$\mathcal{B} \approx \bothcorrelatedBF$ {\bf against} this model.}\\[15pt]
		\hline
		
		\qquad\multirow{5}{4cm}{Slow-spin subpopulation \textit{(dominant component, isotropic, peaked at zero)}} & &
		\multirow{5}{2.2cm}{$\mu_1^a = \mu_2^a = 0$, $\sigma_1^a = \sigma_2^a = \sigma_0$, 
		$\rho^{a,b} = 0$
		} 
		& \multirow{6}{6.5cm}{$\mathcal{B} \approx \BFslowlyspinningtwocomponentdominant$ {\bf in favor} of this model
		compared to two uncorrelated components (above). Similar to the \texttt{BetaSpike-Galaudage} analysis of \cite{Callister:2022qwb}, except there $\sigma_0 \leq 0.1$ which is a disfavored regime, and their bulk spins are identically distributed.
		}  \\[5pt]
		 &    
		\multirow{1}{*}{\resizebox{1cm}{1cm}{\tikzset{every picture/.style={line width=0.75pt}} 

\begin{tikzpicture}[x=0.75pt,y=0.75pt,yscale=-1,xscale=1]

\draw  [color={rgb, 255:red, 248; green, 0; blue, 0 }  ,draw opacity=1 ][fill={rgb, 255:red, 238; green, 0; blue, 0 }  ,fill opacity=0.23 ][line width=1.5]  (31.42,256.81) .. controls (31.42,237.03) and (48.27,220.99) .. (69.05,220.99) .. controls (89.83,220.99) and (106.68,237.03) .. (106.68,256.81) .. controls (106.68,276.59) and (89.83,292.63) .. (69.05,292.63) .. controls (48.27,292.63) and (31.42,276.59) .. (31.42,256.81) -- cycle ;
\draw  [draw opacity=0][fill={rgb, 255:red, 255; green, 255; blue, 255 }  ,fill opacity=1 ] (30.5,38.36) -- (295.5,38.36) -- (295.5,295.36) -- (30.5,295.36) -- cycle(256.95,76.91) -- (69.05,76.91) -- (69.05,256.81) -- (256.95,256.81) -- cycle ;
\draw   (69.05,76.91) -- (256.95,76.91) -- (256.95,256.81) -- (69.05,256.81) -- cycle ;


\draw  [color={rgb, 255:red, 30; green, 0; blue, 255 }  ,draw opacity=0.58 ][fill={rgb, 255:red, 3; green, 0; blue, 255 }  ,fill opacity=0.13 ][line width=1.5]  (207.99,86.8) .. controls (222.34,86.8) and (233.97,106.02) .. (233.97,129.72) .. controls (233.97,153.42) and (222.34,172.63) .. (207.99,172.63) .. controls (193.65,172.63) and (182.02,153.42) .. (182.02,129.72) .. controls (182.02,106.02) and (193.65,86.8) .. (207.99,86.8) -- cycle ;

\end{tikzpicture}}}
		& & \\[10pt]
	    & & & \\[1pt]
		& & & \\[5pt]
		\hline
		
		\qquad\multirow{7}{4cm}{ Zero spin subpopulation} & \multirow{7}{*}{\resizebox{1cm}{1cm}{\tikzset{every picture/.style={line width=0.75pt}} 

\begin{tikzpicture}[x=0.75pt,y=0.75pt,yscale=-1,xscale=1]

\draw  [color={rgb, 255:red, 255; green, 0; blue, 0 }  ,draw opacity=0 ][fill={rgb, 255:red, 238; green, 0; blue, 0 }  ,fill opacity=0 ][line width=1.5]  (45,211.49) .. controls (45,187.67) and (76.9,168.36) .. (116.25,168.36) .. controls (155.6,168.36) and (187.5,187.67) .. (187.5,211.49) .. controls (187.5,235.3) and (155.6,254.61) .. (116.25,254.61) .. controls (76.9,254.61) and (45,235.3) .. (45,211.49) -- cycle ;
\draw  [draw opacity=0][fill={rgb, 255:red, 255; green, 255; blue, 255 }  ,fill opacity=1 ] (30.5,38.36) -- (295.5,38.36) -- (295.5,295.36) -- (30.5,295.36) -- cycle(256.95,76.91) -- (69.05,76.91) -- (69.05,256.81) -- (256.95,256.81) -- cycle ;
\draw   (69.05,76.91) -- (256.95,76.91) -- (256.95,256.81) -- (69.05,256.81) -- cycle ;

Shape: Circle [id:dp19191231190498081] 
\draw  [color={rgb, 255:red, 255; green, 0; blue, 0 }  ,draw opacity=1 ][fill={rgb, 255:red, 255; green, 0; blue, 0 }  ,fill opacity=0.9 ] (53.4,256.06) .. controls (53.4,247.5) and (60.34,240.56) .. (68.9,240.56) .. controls (77.46,240.56) and (84.4,247.5) .. (84.4,256.06) .. controls (84.4,264.62) and (77.46,271.56) .. (68.9,271.56) .. controls (60.34,271.56) and (53.4,264.62) .. (53.4,256.06) -- cycle ;


\draw  [color={rgb, 255:red, 30; green, 0; blue, 255 }  ,draw opacity=0.58 ][fill={rgb, 255:red, 3; green, 0; blue, 255 }  ,fill opacity=0.13 ][line width=1.5]  (184.78,165.81) .. controls (172.71,158.06) and (173.32,135.6) .. (186.14,115.67) .. controls (198.96,95.73) and (219.13,85.86) .. (231.2,93.62) .. controls (243.27,101.38) and (242.66,123.83) .. (229.84,143.76) .. controls (217.02,163.7) and (196.85,173.57) .. (184.78,165.81) -- cycle ;

\end{tikzpicture}}} &\multirow{7}{2.2cm}{$\mu_1^a = \mu_2^a = 0$, $\sigma_1^a = \sigma_2^a = 0$ 
		}& 
		\multirow{7}{6.5cm}{$\mathcal{B} \approx \zerospinBFDominant$ {\bf against} this model if dominant, and $\mathcal{B} \approx \zerospinBFSubDominant$ {\bf against} this model if subdominant (swap components $a \leftrightarrow b$), compared to the full two component model. Similar to the \cite{Tong2022} \texttt{NONIDENTICAL} model and \cite{Mould:2022xeu} \texttt{NONIDENTICAL + ZEROS} model, except here the bulk has correlations.}  
		 \\[25pt]
		& & &
		\\
		& & & \\[25pt]
		\hline
		\hline
		\multirow{1}{*}{  {\bf One Component}
		} &    
		\multirow{3}{*}{\resizebox{2.2cm}{2.2cm}{\tikzset{every picture/.style={line width=0.75pt}} 

\begin{tikzpicture}[x=0.75pt,y=0.75pt,yscale=-1,xscale=1]

\draw  [color={rgb, 255:red, 248; green, 0; blue, 0 }  ,draw opacity=1 ][fill={rgb, 255:red, 238; green, 0; blue, 0 }  ,fill opacity=0.13 ][line width=1.5]  (74.76,116.78) .. controls (96.39,94.74) and (143.7,106.09) .. (180.43,142.14) .. controls (217.16,178.18) and (229.4,225.27) .. (207.77,247.31) .. controls (186.14,269.35) and (138.83,257.99) .. (102.1,221.95) .. controls (65.37,185.91) and (53.13,138.82) .. (74.76,116.78) -- cycle ;
\draw  [draw opacity=0][fill={rgb, 255:red, 255; green, 255; blue, 255 }  ,fill opacity=1 ] (30.5,38.36) -- (295.5,38.36) -- (295.5,295.36) -- (30.5,295.36) -- cycle(256.95,76.91) -- (69.05,76.91) -- (69.05,256.81) -- (256.95,256.81) -- cycle ;
\draw   (69.05,76.91) -- (256.95,76.91) -- (256.95,256.81) -- (69.05,256.81) -- cycle ;

\end{tikzpicture}}}
		&  & \\ 
		& & $\mu_i \sim U(0,1)$ & Mean parameter of the truncated normal \\
		$p(\boldsymbol{\chi}) = N_{[\boldsymbol{0},\boldsymbol{1}]}\left(\boldsymbol{\chi}\ | \boldsymbol{\mu}, \boldsymbol{\Sigma}\right)\,$ & & $\sigma_i \sim U(0,1)$ & Scale parameters of the truncated normal \\[4pt]
		\multirow{1}{4cm}{$\mathcal{B} \approx \BFsecondcomponentexists$ {\bf against} this model compared to the two component model.}
		& & $\rho \sim U(-1,1)$ & Correlation parameter of the truncated normal \\
        & & &  \\
		& & & \\[2pt]
		\hline
		\hline
		&&& \\[-5pt]
		\qquad \multirow{2}{4cm}{Primary and secondary uncorrelated} &
		\multirow{2}{*}{\resizebox{1cm}{1cm}{\tikzset{every picture/.style={line width=0.75pt}} 

\begin{tikzpicture}[x=0.75pt,y=0.75pt,yscale=-1,xscale=1]

\draw  [color={rgb, 255:red, 248; green, 0; blue, 0 }  ,draw opacity=1 ][fill={rgb, 255:red, 238; green, 0; blue, 0 }  ,fill opacity=0.13 ][line width=1.5]  (162.69,87.08) .. controls (189.13,86.98) and (210.7,122.62) .. (210.87,166.68) .. controls (211.04,210.74) and (189.74,246.54) .. (163.31,246.64) .. controls (136.87,246.74) and (115.3,211.1) .. (115.13,167.04) .. controls (114.96,122.98) and (136.26,87.18) .. (162.69,87.08) -- cycle ;
\draw  [draw opacity=0][fill={rgb, 255:red, 255; green, 255; blue, 255 }  ,fill opacity=1 ] (30.5,38.36) -- (295.5,38.36) -- (295.5,295.36) -- (30.5,295.36) -- cycle(256.95,76.91) -- (69.05,76.91) -- (69.05,256.81) -- (256.95,256.81) -- cycle ;
\draw   (69.05,76.91) -- (256.95,76.91) -- (256.95,256.81) -- (69.05,256.81) -- cycle ;

\end{tikzpicture}}} & $\hat{\rho} = 0$ & \multirow{3}{6.5cm}{$\mathcal{B} \approx \onepopuncorrelatedBF$ {\bf against} this model when compared to the one component model. \\ Similar to the \cite{Mould:2022xeu} \texttt{Nonidentical} model.} \\[15pt]
		& & & \\[5pt]
		\hline
		&&& \\[-5pt]
		\multirow{3}{*}{\qquad Primary and secondary are IID} &
		\multirow{3}{*}{\resizebox{1cm}{1cm}{\tikzset{every picture/.style={line width=0.75pt}} 

\begin{tikzpicture}[x=0.75pt,y=0.75pt,yscale=-1,xscale=1]

\draw  [color={rgb, 255:red, 248; green, 0; blue, 0 }  ,draw opacity=1 ][fill={rgb, 255:red, 238; green, 0; blue, 0 }  ,fill opacity=0.13 ][line width=1.5]  (162.78,109.17) .. controls (196.44,109.04) and (223.83,134.77) .. (223.95,166.63) .. controls (224.08,198.49) and (196.89,224.42) .. (163.22,224.55) .. controls (129.56,224.68) and (102.17,198.95) .. (102.05,167.09) .. controls (101.92,135.23) and (129.11,109.3) .. (162.78,109.17) -- cycle ;
\draw  [draw opacity=0][fill={rgb, 255:red, 255; green, 255; blue, 255 }  ,fill opacity=1 ] (30.5,38.36) -- (295.5,38.36) -- (295.5,295.36) -- (30.5,295.36) -- cycle(256.95,76.91) -- (69.05,76.91) -- (69.05,256.81) -- (256.95,256.81) -- cycle ;
\draw   (69.05,76.91) -- (256.95,76.91) -- (256.95,256.81) -- (69.05,256.81) -- cycle ;

\end{tikzpicture}}} & $\hat{\rho}$, $\mu_1 = \mu_2$ &  \multirow{3}{6.5cm}{$\mathcal{B} \approx  \onepopuncorrelatedIIDBF$ {\bf in favor} this model when compared to the one component model. \\ This model is similar to the IID models often used, for example by \cite{KAGRA:2021duu} in their fiducial model.} \\
		& & $\sigma_1 = \sigma_2$ & \\[31pt]
		\hline
		\multirow{4}{*}{\qquad All secondaries not spinning} &
		\multirow{4}{*}{\resizebox{1cm}{1cm}{\tikzset{every picture/.style={line width=0.75pt}} 

\begin{tikzpicture}[x=0.75pt,y=0.75pt,yscale=-1,xscale=1]

\draw  [draw opacity=0][fill={rgb, 255:red, 255; green, 255; blue, 255 }  ,fill opacity=1 ] (30.5,38.36) -- (295.5,38.36) -- (295.5,295.36) -- (30.5,295.36) -- cycle(256.95,76.91) -- (69.05,76.91) -- (69.05,256.81) -- (256.95,256.81) -- cycle ;
\draw   (69.05,76.91) -- (256.95,76.91) -- (256.95,256.81) -- (69.05,256.81) -- cycle ;

\draw  [color={rgb, 255:red, 30; green, 0; blue, 255 }  ,draw opacity=0 ][fill={rgb, 255:red, 3; green, 0; blue, 255 }  ,fill opacity=0 ][line width=1.5]  (241.99,61.2) .. controls (256.34,61.2) and (267.97,80.42) .. (267.97,104.12) .. controls (267.97,127.82) and (256.34,147.03) .. (241.99,147.03) .. controls (227.65,147.03) and (216.02,127.82) .. (216.02,104.12) .. controls (216.02,80.42) and (227.65,61.2) .. (241.99,61.2) -- cycle ;
\draw  [color={rgb, 255:red, 248; green, 0; blue, 0 }  ,draw opacity=1 ][fill={rgb, 255:red, 238; green, 0; blue, 0 }  ,fill opacity=0.13 ][line width=1.5]  (155,253.87) .. controls (202.82,253.69) and (241.59,254.54) .. (241.59,255.77) .. controls (241.6,257) and (202.84,258.15) .. (155.02,258.33) .. controls (107.2,258.52) and (68.44,257.67) .. (68.43,256.44) .. controls (68.43,255.2) and (107.19,254.06) .. (155,253.87) -- cycle ;

\end{tikzpicture}}} & \multirow{4}{*}{$\mu_2 = \sigma_2 = 0$}  &\multirow{4}{6.5cm}{$\mathcal{B} \approx \allsecondarieszeroBF$ {\bf in favor} of all secondaries nonspinning over the two component model. ($\mathcal{B} \approx \allsecondarieszeroBFoveronecomponent$ if compared to the one component model)} \\
		& & & \\[25pt]
		\hline
	\end{tabular}
	\caption{Summary of two-component and one-component models, their priors, and limiting cases. We include the \acp{BF} computed using the \ac{SDDR} for each sub-model, compared to either the full two-component model with correlations or the one-component model (for the second section). Additionally, we note if some cases resemble those explored in the literature, modulo interchange of a beta distribution with a truncated normal.}
	\label{tab:model-summary}
\end{table*}

To allow us to probe the population properties of astrophysical BBHs, we start by using the fiducial mass, redshift and spin orientation models as defined in \cite{KAGRA:2021duu}. More specifically we use the \texttt{PowerLawPlusPeak} model for the masses, \texttt{PowerLaw} model for redshift and we use only the part of the \texttt{Default} spin model that pertains to the spin orientation given by,
\begin{equation}
	p\left(\bm{z}_t \mid \zeta, \sigma_t\right)=\zeta N_{[\boldsymbol{0},\boldsymbol{1}]}\left(\bm{z}_t \mid \sigma_t \bm{I}_2 \right)+(1-\zeta)\frac{1}{4}\,,
	\label{eq:orientationmodel}
\end{equation}
where $z_{t,i} = \cos \theta_i$ and $\theta_i$ is the tilt angle between each spin and the orbital angular momentum of the binary. 
In other words, the cosine tilts are either drawn independently from a half-normal with scale parameter $\sigma_t$ (aligned) or from a uniform distribution (isotropic), with the fraction of aligned spin binaries given by $\zeta$.
The spin magnitude model is given in Eq.~\eqref{eq:twocomponentmodel} and the priors on the population parameters for it are given in Table~\ref{tab:model-summary}. 
The priors for the rest of the population parameters are set as the same as those used in \cite{KAGRA:2021duu}, with the exception of the width of the isotropic spin distribution, where we allow the prior to go all the way to zero [$\sigma_t \sim U(0,4)$, as opposed to $\sigma_t \sim U(0.1,4)$].

We use truncated normal distributions to model the spin magnitudes as opposed to beta distributions, since the truncated normal distribution is better at recovering sharp distributions near the edges (e.g.~a sharp peak at $\chi \approx 0$) than a beta distribution. 
In addition, a beta distribution may become singular for certain parameter values ($\alpha, \beta > 1$), causing peaks at both $\chi \to 0$ and $\chi \to 1$ if these are allowed. 
When using a hyper-prior to remove these regions the resulting prior predictive distribution is not flat in the spin magnitudes, i.e., it gives some preference to $\chi \approx 0.5$ over $\chi \approx 0$ or $\chi \approx 1$.
Our hyper-priors result in very nearly flat prior predictive distributions over the spin magnitudes.

\subsection{Summary of TGMM population analysis method}
Here we briefly summarize the \ac{TGMM} population analysis method. 
We refer to~\cite{TGMM_Methods} for the full details of the method. 
As part of a hierarchical Bayesian inference procedure on \ac{GW} data \citep{Mandel:2018mve,Thrane:2018qnx,Vitale:2020aaz}, we need to efficiently estimate certain marginalized likelihoods using importance sampling. 
This requires samples from individual-event posteriors (with prior weights) and samples from an injection campaign to estimate detection sensitivity in different parts of parameter space.
Those ingredients can allow us to infer the distribution of the population parameters in the presence of selection effects.

In general, importance sampling is needed to estimate integrals of the form
\begin{equation}
	\label{eq:generic-integral}
	I(\boldsymbol{\Lambda}) = \int p(\boldsymbol{\theta} | \boldsymbol{\Lambda}) \frac{p(\boldsymbol{\theta} \mid \cdot\, )}{W(\boldsymbol{\theta})} \mathrm{d}\boldsymbol{\theta} 
	\approx \left\langle   \frac{p(\boldsymbol{\theta} \mid \boldsymbol{\Lambda})}{W(\boldsymbol{\theta})}  \right\rangle_{\boldsymbol{\theta} \sim p(\boldsymbol{\theta}\mid \cdot)}\, ,
\end{equation} 
when we have samples from a distribution $p(\boldsymbol{\theta}\mid \cdot\,)$ conditional on some assumptions or observations (represented by ``$\cdot$''), as well as access to sampling or prior weights $W(\boldsymbol{\theta})$. 
In particular, the posterior over the population parameters using $N$ events is given by
\begin{equation}
	p(\boldsymbol{\Lambda} \mid \{d_i\}) = \pi(\boldsymbol{\Lambda})\, \xi(\boldsymbol{\Lambda})^{-N}\prod_{i}^N \mathcal{L}(d_i \mid \boldsymbol{\Lambda}) \,,
	\label{eq:populationposterior}
\end{equation}
where $\pi(\boldsymbol{\Lambda})$ is the hyperprior, $\xi(\boldsymbol{\Lambda})$ is the detection efficiency (defined below) and the $\mathcal{L}(d_i \mid \boldsymbol{\Lambda})$ are the individual event-level marginalized likelihoods, given by
\begin{equation}
	\mathcal{L}(d_i \mid \boldsymbol{\Lambda})=\int  p(\boldsymbol{\theta} \mid \boldsymbol{\Lambda}) \frac{p(\boldsymbol{\theta} \mid d_i)}{\pi(\boldsymbol{\theta} \mid \emptyset)} \mathrm{d}\boldsymbol{\theta}
	\approx \left\langle   \frac{p(\boldsymbol{\theta} \mid \boldsymbol{\Lambda})}{\pi(\boldsymbol{\theta} \mid \emptyset)}  
	\right\rangle_{\boldsymbol{\theta} \sim p(\boldsymbol{\theta} \mid d_i)}\,,
	\label{eq:eventpoplikelihood}
\end{equation}
where in turn $p(\boldsymbol{\theta} \mid d_i)$ is the posterior for event $i$, and $\pi(\boldsymbol{\theta} \mid \emptyset)$ is the sampling prior used in the original Bayesian inference for that analysis, and $p(\boldsymbol{\theta} \mid \boldsymbol{\Lambda})$ is the population model whose parameters we wish to infer. 
The population-averaged detection efficiency is calculated using
\begin{equation}
	\xi(\boldsymbol{\Lambda}) =  \int p(\boldsymbol{\theta} \mid \boldsymbol{\Lambda})  P_{\rm det}(\boldsymbol{\theta})\, \mathrm{d}\boldsymbol{\theta} \approx  \int p(\boldsymbol{\theta} \mid \boldsymbol{\Lambda_0})\,  P_{\rm det}(\boldsymbol{\theta}) \frac{p(\boldsymbol{\theta} \mid \boldsymbol{\Lambda})}{p(\boldsymbol{\theta} \mid \boldsymbol{\Lambda}_0)} \mathrm{d}\boldsymbol{\theta} 
	\approx
	\frac{N_{\rm det}}{N_{\rm draw}} \left\langle \frac{p(\boldsymbol{\theta} \mid \boldsymbol{\Lambda})}{p(\boldsymbol{\theta} \mid \boldsymbol{\Lambda}_0)}  \right\rangle_{\boldsymbol{\theta} \sim p_{\rm det}(\boldsymbol{\theta} | \boldsymbol{\Lambda}_0)}\, ,
	\label{eq:detectionefficiency}
\end{equation}
where $P_{\rm det}(\boldsymbol{\theta})$ is the probability of detecting a signal with parameters $\boldsymbol{\theta}$, $\boldsymbol{\Lambda}_0$ represents some fiducial population from which $N_{\rm draw}$ signals are simulated to estimate detection efficiency and obtain $N_{\rm det}$ samples of detected signals from $p_{\rm det}(\boldsymbol{\theta} \mid \boldsymbol{\Lambda}_0)$, which is proportional, but not equal, to the fiducial population density times the detection probability, i.e., $p(\boldsymbol{\theta\mid\boldsymbol{\Lambda}_0}) \propto p(\boldsymbol{\theta}\mid \boldsymbol{\Lambda_0})\, \mathbf{P}_{\rm det}(\boldsymbol{\theta})$ with proportionality constant $N_{\rm det} / N_{\rm draw}$.
We can see that Eqs.~\eqref{eq:detectionefficiency} and \eqref{eq:eventpoplikelihood} are special cases of Eq.~\eqref{eq:generic-integral}.

For an integral of the form of Eq.~\eqref{eq:generic-integral}, we can take the samples from $p(\boldsymbol{\theta}\mid \cdot\,)$, representing the event posteriors or the detected injection set, and fit to them a weighted sum of \textit{truncated} multivariate Gaussians 
\begin{equation} \label{eq:tgmm}
	p(\boldsymbol{\theta}\mid \cdot\,) = \sum_k w_k N_{[{\bf a},{\bf b}]}(\boldsymbol{\theta} \mid \boldsymbol{\mu}_k, {\bf \Sigma}_k) \,,
\end{equation}
where $w_k$ are mixture weights, and $\boldsymbol{a}$ and $\boldsymbol{b}$ are the two bounding corners of the hypercube defining the limits of the parameters $\boldsymbol{\theta}$. 
The fitting procedure is described in \cite{Lee2012}, with improvements drawn from \cite{gmm-anealing, weighted-samples, expectation-conjugate-gradient, TGMM_Methods}. 
We have released a package to perform this fitting procedure, \texttt{truncatedgaussianmixtures} \citep{truncatedgaussianmixtures}.

Now we assume that the population model is separable, such that the mass and redshift sector (with parameters denoted $\boldsymbol{\theta}^{m,z}$) separate from the spin sector (covering both spin magnitudes and spin tilts, denoted $\boldsymbol{\theta}^\chi$),
\begin{equation}
	p(\boldsymbol{\theta} \mid \boldsymbol{\Lambda}) = p(\boldsymbol{\theta}^\chi \mid \boldsymbol{\Lambda}^\chi)\, p(\boldsymbol{\theta}^{m,z} \mid \boldsymbol{\Lambda}^{m,z}) \,,
		\label{eq:seperatedmodel}
\end{equation}
and that, in its most general form, the population model in the spin sector is some mixture of truncated multivariate Gaussians (or uniform distributions). 
For example, our population model for the spin magnitude and spin orientation is of this form, as seen in Eqs.~\eqref{eq:twocomponentmodel} and~\eqref{eq:orientationmodel}.
In general, we can allow for any weighted sum of such separable sub-models using our methods.

To leverage this separability, while fitting the \acp{TGMM} of Eq.~\eqref{eq:tgmm} 
we require that the covariance matrix for each component does not create correlations between the spin sector and the other sectors, i.e.,
\begin{equation}
	p(\boldsymbol{\theta} \mid \cdot\, ) = \sum_k w_k N_{[{\bf a},{\bf b}]}(\boldsymbol{\theta}^\chi \mid \boldsymbol{\mu}^{\chi}_k, {\bf \Sigma}^{\chi}_k)\, N_{[{\bf a},{\bf b}]}(\boldsymbol{\theta}^{m,z} \mid \boldsymbol{\mu}^{m,z}_k, {\bf \Sigma}^{m,z}_k) \,.
	\label{eq:seperatedtgmmfit}
\end{equation}
This does not impose strong restrictions in the distributions that can be fit and, in particular, does not imply that we cannot capture correlations across the two sectors: although each individual \ac{TGMM} component cannot have cross-sector correlations across, cross-sector correlation structure can still be captured by the joint arrangement of multiple \ac{TGMM} components. 
Indeed, the action of several uncorrelated components together can construct correlations (e.g., as an extreme case, \acp{KDE} with uncorrelated bandwidth matrices can easily represent distributions with large-scale covariances).

By substituting Eqs.~\eqref{eq:seperatedmodel} and \eqref{eq:seperatedtgmmfit} into \eqref{eq:generic-integral} with $W(\boldsymbol{\theta})$ set to the sampling prior for concreteness (equivalently, the sampling distribution for sensitivity injections), we get,
\begin{equation}
	I(\boldsymbol{\Lambda}) = \sum_k w_k
\int N_{[{\bf a},{\bf b}]}(\boldsymbol{\theta}^\chi\mid \boldsymbol{\mu}^{\chi}_k, {\bf \Sigma}^{\chi}_k)\, p(\boldsymbol{\theta}^\chi \mid \boldsymbol{\Lambda}^\chi)\, \mathrm{d}\boldsymbol{\theta}^\chi 
\int N_{[{\bf a},{\bf b}]}(\boldsymbol{\theta}^{m,z} \mid \boldsymbol{\mu}^{m,z}_k, {\bf \Sigma}^{m,z}_k)\, \frac{p(\boldsymbol{\theta}^{m,z} \mid \boldsymbol{\Lambda}^{m,z})}{\pi(\boldsymbol{\theta}^{m,z} \mid \emptyset)}\,  \mathrm{d}\boldsymbol{\theta}^{m,z} \,,
	\label{eq:seperatedintegral-0}
\end{equation}
where we have assumed for now that the sampling prior in spin space is flat, i.e., $\pi(\boldsymbol{\theta}^\chi \mid \emptyset) = 1$,
which is true of standard \ac{LVK} priors on the spin magnitudes and cosines of the spin tilts (this constraint is removed in the full description of our methodology in \citealt{TGMM_Methods}). 
We then rewrite the above as
\begin{equation}
		I(\boldsymbol{\Lambda}) = \sum_k w_k\,
		I(\boldsymbol{\mu}^{\chi}_k, {\bf \Sigma}^{\chi}_k, \boldsymbol{\Lambda}_\chi)
	\left\langle \frac{p(\boldsymbol{\theta}^{m,z} \mid \boldsymbol{\Lambda}^{m,z})}{\pi(\boldsymbol{\theta}^{m,z} \mid \emptyset)}\right\rangle_{\boldsymbol{\theta}_{m,z} \sim N_{[{\bf a},{\bf b}]}(\boldsymbol{\theta}^{m,z} \mid \boldsymbol{\mu}^{m,z}_k, {\bf \Sigma}^{m,z}_k)} \,.
		\label{eq:seperatedintegral-1}
\end{equation}
Here the integral $I(\boldsymbol{\mu}^{\chi}_k, {\bf \Sigma}^{\chi}_k, \boldsymbol{\Lambda}^\chi)$ can be handled semi-analytically using properties of Gaussians (using custom numerical routines we implement), while the expectation over the mass and redshift can be approximated using Monte Carlo estimation, as is standard int the \ac{LVK} literature. 
To handle cuts and stabilize the \ac{TGMM} fit in the mass and redshift domains, we note that each \ac{TGMM} fit allows us to extract an assignment of each sample $\boldsymbol{\theta}_i$ to a specific \ac{TGMM} component $k$. 
Armed with this assignment, we can then rewrite the expectation over the mass and redshift as an expectation over the \textit{original} posterior samples that were assigned to each component $k$.
This leaves us with the expression
\begin{equation}
		I(\boldsymbol{\Lambda}) = \sum_k w_k \,
		I(\boldsymbol{\mu}^{\chi}_k, {\bf \Sigma}^{\chi}_k, \boldsymbol{\Lambda}^\chi)
		\left[\frac{1}{N_k}\sum_j^{N_k} \frac{p(\boldsymbol{\theta}_{j,k}^{m,z} \mid \boldsymbol{\Lambda}^{m,z})}{\pi(\boldsymbol{\theta}_{j,k}^{m,z} \mid \emptyset)}\right]\,,
		\label{eq:seperatedintegral-final}
\end{equation}
where $\boldsymbol{\theta}_{j,k}$ is the $j$th sample assigned to the $k$th component.
Equation \eqref{eq:seperatedintegral-final} is efficient to evaluate.
We have published a python package (\texttt{gravpop}) to perform population analysis using this method \citep{gravpop}.

We explore the variance properties of the above estimator \eqref{eq:seperatedintegral-final} for $I(\boldsymbol{\Lambda})$ in the companion paper \cite{TGMM_Methods}, and find no singular behaviour of the estimator's variance as population features become narrow. 
Since we only perform the Monte Carlo estimate over the mass and redshift sectors, the variance of this estimator is lower than in analyses using Monte Carlo methods across all sectors. 
We have not computed the variance of the population likelihood estimators explicitly in this study, nor applied any cuts associated with the variance of the estimator at given hyper-parameter values (e.g., the data-dependent priors discussed in \citealt{KAGRA:2021duu}).

\subsection{Details on the computation of SDDR}
\label{sec:SDDRMethods}
To compute a \ac{SDDR}, we need an estimate of the posterior density at a point of interest that often lies at the edges of our domain (e.g., $\chi = 0$). 
We use a custom multivariate \ac{KDE} to get an estimate of the marginalized hyper-posterior from samples. 
We describe this multivariate \ac{KDE} in \cite{TGMM_Methods} but note here that it does not impose a zero derivative at the edge and has no bias at the boundary at $\mathcal O(b^0)$, where $b$ is the kernel bandwidth, an issue that often plagues other \ac{KDE} techniques and makes them unsuitable for \acp{SDDR}.
Since the bias in our method is $\mathcal{O}(b)$, we can achieve a better estimate of the density at the edge of parameter space by increasing the number of samples (reducing the bandwidth).
Briefly, our \ac{KDE} maps each point to a truncated multivariate normal with some uncorrelated bandwidth vector $\mathbf{b}$ ($b = |\mathbf{b}|$), but the position of this multivariate normal is moved away from the location of the sample in such a way that the overall \ac{KDE} estimate has a bias of only order $\mathcal{O}(b)$. 
This is similar to what can be achieved using \acp{KDE} with reflective boundary conditions, such as those used by \cite{Callister:2022qwb} to compute \acp{SDDR}.
A benefit of our method as compared to reflective \acp{KDE} is that we do not require the derivative of the kernel be zero at the boundaries, and we do not need the additional kernels to enforce reflective boundaries (which can require a large number of additional kernels in higher dimensions).

We highlight our procedure of extracting \acp{BF} and their associated bootstrap uncertainties using the \ac{SDDR} with an illustrative example. 
Consider computing the \ac{SDDR} for our fiducial spin-magnitude model, which supports two subpopulations, in the limit that all the secondary \acp{BH} in the dominant population are nonspinning. 
This corresponds to $\mu^a_2 \to 0$  and $\sigma^a_2 \to 0$. 
We first draw $N$ samples with replacement  from our hyper-posterior samples. 
In this two-dimensional space we then get a \ac{KDE} estimate of the marginal hyper-posterior over $\mu^a_2 $ and $\sigma^a_2$, $\hat{p}(\mu^a_2,\sigma^a_2)$. Then we evaluate the estimator for the \ac{BF}, $\hat{\mathcal{B}}$, given by
\begin{equation}
	 \hat{\mathcal B} = \frac{\hat{p}(\mu^a_2=0,\sigma^a_2=0)}{\pi(\mu^a_2=0,\sigma^a_2=0)}\,,
\end{equation}
where $\pi(\mu^a_2,\sigma^a_2)$ is the marginalized hyper-prior for our analysis. 
This gives us an estimate of the \ac{BF} from one sampling of our hyper-posterior. 
We subsequently repeat this process $\mathcal{O}(100)$ times, and get a series of \ac{BF} estimates. 
With these estimates we compute the bootstrapped median and $90\%$ highest-confidence interval for our estimate of the \ac{BF}.

\subsection{Details on the computation of BF for event assignments}
\label{sec:Appxeventassignments}
We use the following method to compute the \ac{BF} for a given event in favor of the hypothesis that the even belongs to a given of a particular subpopulation, which is used in Fig.~\ref{fig:OrientationMagnitudeMassTrend}.
Assume that the population model breaks into two subpopulations, subpopulation $A$ and subpopulation $B$, with the fraction $\eta$ of the population in $A$. 
The evidence of some event $i$ under the population prior described by $\boldsymbol{\Lambda}$ is given by $Z_i(\boldsymbol{\Lambda}) = \mathcal{L}(d_i | \boldsymbol{\Lambda})$ as defined in \eqref{eq:eventpoplikelihood}. 

We take our hyper-posterior samples, and for each  sample $
\boldsymbol{\Lambda}_j$, we compute the evidence $Z_i(\boldsymbol{\Lambda}_j^{\eta = 1})$ under the hypothesis that only subpopulation $A$ exists, $\eta \to 1$, and then compute the evidence $Z_i(\boldsymbol{\Lambda}_j^{\eta = 0})$ under the hypothesis that only subpopulation $B$ exists, $\eta \to 0$. 
Here $\boldsymbol{\Lambda}_j^{\eta = 0}$ simply means we set $\eta = 0$ for that sample.
The ratio gives our estimate of the \ac{BF} between the two population hypotheses for event $i$
\begin{equation}
	\mathcal{B}_{i,j} = \frac{Z_i(\boldsymbol{\Lambda}_j^{\eta = 1})}{Z_i(\boldsymbol{\Lambda}_j^{\eta = 0})}\,,
\end{equation} 
giving us our posterior over \acp{BF} for event $i$. 
We use these to compute the expected $\log_{10}\mathcal B_i$ from our posterior over $\mathcal{B}_i$ using the samples $\mathcal B_{i,j}$.
As with our SDDR estimates of the \acp{BF}, sec.~\ref{sec:SDDRMethods}, we report the mean and compute the $90\%$ confidence interval of our \acp{BF} estimates using bootstrap.
 We report the mean  in Fig.~\ref{fig:OrientationMagnitudeMassTrend} for the spin-orientation subpopulations and the spin-magnitude subpopulations.
 The $90\%$ confidence interval of these estimates is relatively small, with a maximum width of $\pm 0.2$ for $\langle \log_{10}\mathcal B_{\text{rapid spin}}\rangle$ and $\pm 0.03$ for $\langle \log_{10}\mathcal B_{\text{aligned}}\rangle$ over all events.

\section{Further results: Corner plots}
\label{sec:ResultsAppx}

\begin{figure}[tb]
	\includegraphics[width=\textwidth{}]{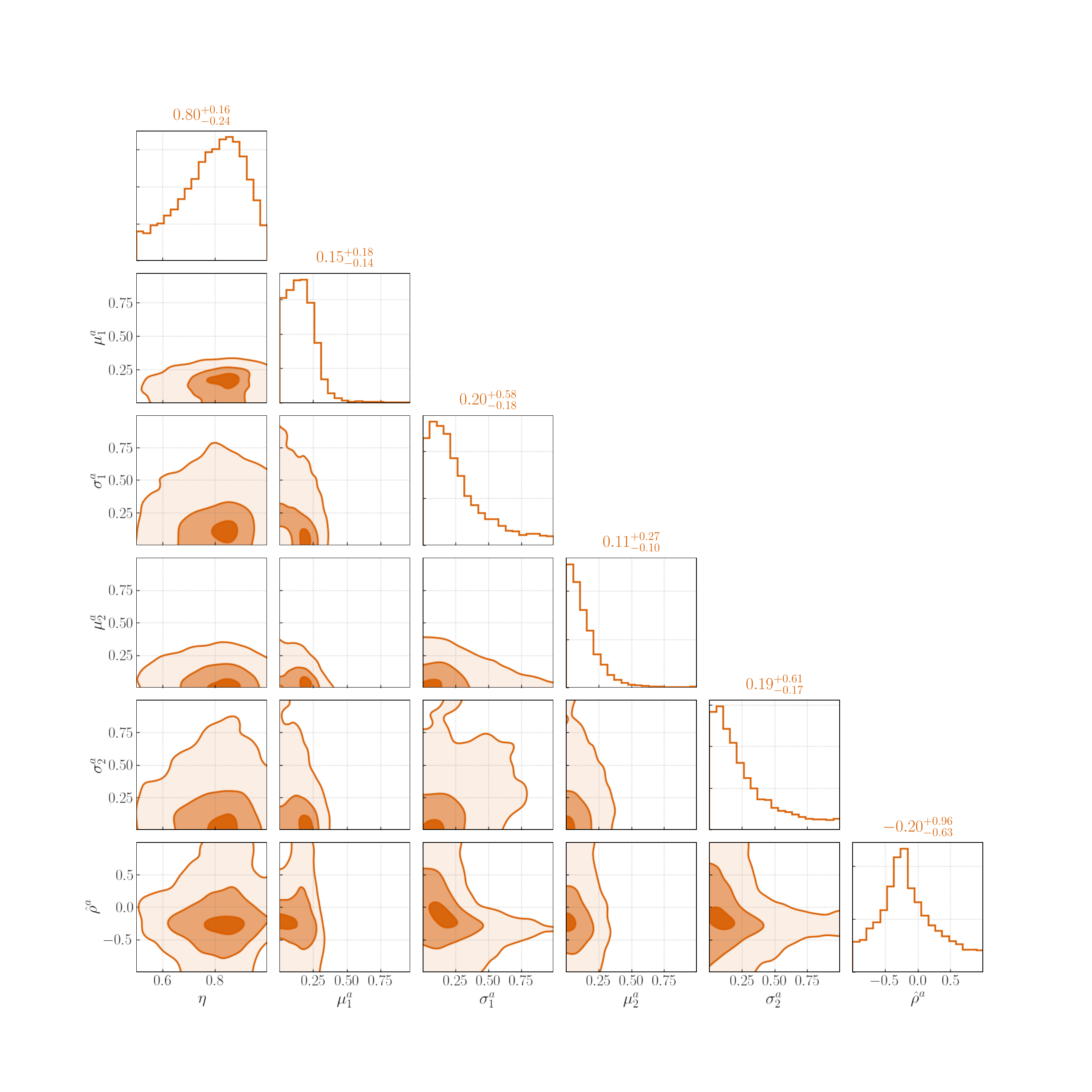}
	\caption{
		Corner plot for the posteriors of the hyper-parameters pertaining to the dominant subpopulation.
	}
	\label{fig:CornerComponent1}
\end{figure}

\begin{figure}[tb]
 	\includegraphics[width=\textwidth{}]{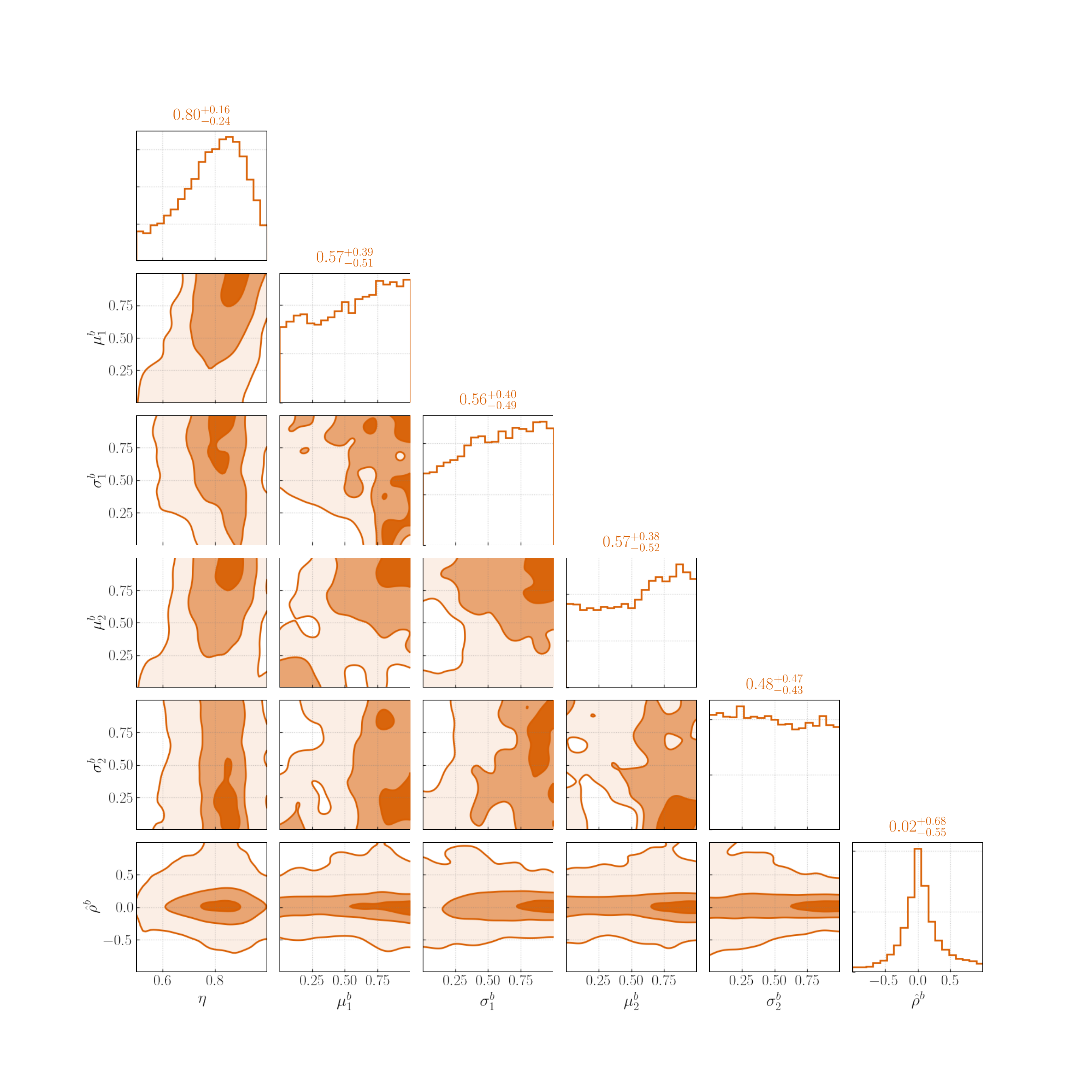}
 	\caption{
 		Corner plot for the posteriors of the hyper-parameters pertaining to the subdominant subpopulation.
 	}
 	\label{fig:CornerComponent2}
 \end{figure}

In this Appendix we show corner plots for the parameters governing each subpopulation in our fiducial two-population spin-magnitude analysis.
Figure~\ref{fig:CornerComponent1} shows the two-dimensional and one-dimensional marginals for the parameters of the dominant, slowly spinning population.
Figure~\ref{fig:CornerComponent2} shows the same marginals but for the parameters of the subdominant, rapidly spinning population.
As compared to the parameters of the dominant population, the parameters of this component are less informed.
The parameters controlling the mean $\mu^b_i$ display the bi-modality discussed above, with a preferred mode at large $\mu^b_1$--$\mu^b_2$ values and a less significant mode at small $\mu^b_1$--$\mu^b_2$ values.
This second mode is due to residual degeneracy between the two subpopulations, with these small values possible only when the fraction $\eta$ is close to $1/2$.

\begin{figure}[tb]
	\includegraphics[width=\textwidth{}]{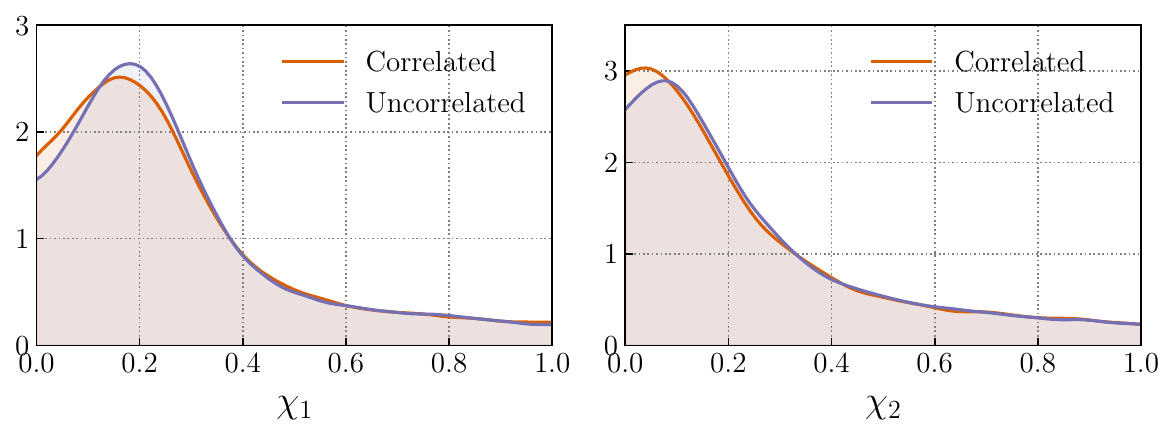}
	\caption{
		Marginals over the \ac{PPD} of the primary and secondary spin magnitude $\chi_1$ and $\chi_2$. We compare the analysis allowing sub-populations to have correlations between $\chi_1$ and $\chi_2$, to one where they are not correlated. One can see that the introduction of correlations to the model does not leave a very significant imprint on the marignal distributions, but can be seen much more clearly on the two dimensional \ac{PPD} (Fig.~\ref{fig:TwoComponentPosteriorPredictive}). 
	}
	\label{fig:PPD_marginals}
\end{figure}

Figure~\ref{fig:PPD_marginals} shows the \acp{PPD} of the individual spins from our fiducial two-population analysis, with and without correlation in the component covariances.
The existence of a highly spinning subpopulation is difficult to infer from these individual marginal \acp{PPD}, appearing only as a heavier tail in the $\chi_1$ \ac{PPD} when correlation is included.

\bibliographystyle{aasjournal}
\bibliography{Spin_Pop_Letter.bbl}

\end{document}